\documentclass{aa}  

\usepackage{graphicx}
\usepackage{txfonts}
\usepackage{hyperref}
\def\zlens{z_{\mathrm{l}}}
\def\zsource{z_{\mathrm{s}}}

\def\muh{\mu_{\mathrm{h}}}
\def\muhzero{\mu_{\mathrm{h},0}}
\def\zetah{\zeta_{\mathrm{h}}}
\def\lambdah{\lambda_{\mathrm{h}}}
\def\gammah{\gamma_{\mathrm{h}}}
\def\betah{\beta_{\mathrm{h}}}
\def\xih{\xi_{\mathrm{h}}}
\def\nuh{\nu_{\mathrm{h}}}
\def\sigmah{\sigma_{\mathrm{h}}}

\def\muR{\mu_{\mathrm{R}}}
\def\muRzero{\mu_{\mathrm{R},0}}
\def\zetaR{\zeta_{\mathrm{R}}}
\def\betaR{\beta_{\mathrm{R}}}
\def\sigmaR{\sigma_{\mathrm{R}}}

\def\nuC{\nu_{\mathrm{C}}}
\def\muC{\mu_{\mathrm{C}}}
\def\muCzero{\mu_{\mathrm{C},0}}
\def\lambdaC{\lambda_{\mathrm{C}}}
\def\gammaC{\gamma_{\mathrm{C}}}
\def\zetaC{\zeta_{\mathrm{C}}}
\def\betaC{\beta_{\mathrm{C}}}
\def\xiC{\xi_{\mathrm{C}}}
\def\sigmaC{\sigma_{\mathrm{C}}}

\def\sigmacr{\Sigma_{\mathrm{cr}}}

\def\mstar{M_*}

\def\mhalo{M_{\mathrm{h}}}

\def\rhalo{r_{\mathrm{h}}}
\def\chalo{c_{\mathrm{h}}}

\def\reff{R_{\mathrm{e}}}

\def\mhalo{M_{\mathrm{h}}}

\def\ewha{\mathrm{EW}_{\mathrm{H}\alpha}}

\def\lumr{L_r}

\def\slope{\Gamma_{10}}

\def\lten{L_{r,10}}

\def\epsobs{\epsilon_t^{(\mathrm{obs})}}

\def\compell{\boldsymbol\epsilon}
\def\ellone{\epsilon_1}
\def\elltwo{\epsilon_2}
\def\compellt{\epsilon_t}
\def\redsheart{g_t}

\def\hyperp{\boldsymbol{\eta}}
\def\hypermap{\boldsymbol{\eta}^{(\mathrm{MAP})}}

\def\data{\mathbf{d}}
\def\datai{\mathbf{d}_i}

\def\Sref#1{Section~\ref{#1}\xspace}
\def\Fref#1{Figure~\ref{#1}\xspace}
\def\Tref#1{Table~\ref{#1}\xspace}
\def\Eref#1{Equation~\ref{#1}\xspace}

\def\pr{{\rm P}}

\begin{document} 

   \title{The dark matter halo masses of elliptical galaxies as a function of observationally robust quantities}
   \titlerunning{The dark matter halos of elliptical galaxies}
   \authorrunning{Sonnenfeld et al.}


   \author{Alessandro Sonnenfeld\inst{\ref{leiden}}\and
           Crescenzo Tortora\inst{\ref{inaf-na}}\and
           Henk Hoekstra\inst{\ref{leiden}}\and
           Marika Asgari\inst{\ref{edinburgh},\ref{hull}}\and
           Maciej Bilicki\inst{\ref{warsaw}}\and
           Catherine Heymans\inst{\ref{edinburgh},\ref{bochum}}\and
           Hendrik Hildebrandt\inst{\ref{bochum}}\and
           Konrad Kuijken\inst{\ref{leiden}}\and
           Nicola R. Napolitano\inst{\ref{zhuhai},\ref{csst}}\and
           Nivya Roy\inst{\ref{kerala}}\and
           Edwin Valentijn\inst{\ref{groningen}}\and
           Angus H. Wright\inst{\ref{bochum}}
          }

   \institute{Leiden Observatory, Leiden University, PO Box 9513, NL-2300 RA Leiden, the Netherlands\\
              \email{sonnenfeld@strw.leidenuniv.nl}\label{leiden}
             \and
            INAF -- Osservatorio Astronomico di Capodimonte, Salita Moiariello 16, 80131, Naples, Italy\label{inaf-na}\and
            Institute for Astronomy, University of Edinburgh, Royal Observatory, Blackford Hill, Edinburgh, EH9 3HJ, UK\label{edinburgh}\and
            E. A. Milne Centre, University of Hull, Cottingham Road, Hull, HU6 7RX, UK\label{hull}\and
            Center for Theoretical Physics, Polish Academy of Sciences, al. Lotników 32/46, 02-668 Warsaw, Poland\label{warsaw}\and
            Ruhr University Bochum, Faculty of Physics and Astronomy, Astronomical Institute (AIRUB), German Centre for Cosmological Lensing, 44780 Bochum, Germany\label{bochum}\and
            School of Physics and Astronomy, Sun Yat-sen University, Guangzhou 519082, Zhuhai Campus, P.R. China\label{zhuhai}\and
            CSST Science Center for Guangdong-Hong Kong-Macau Great Bay Area, Zhuhai, China, 519082\label{csst}\and
            Department of Physics, Carmel College, Mala, Thrissur 680732, Kerala, India\label{kerala}\and
            Kapteyn Institute, University of Groningen, PO Box 800, NL 9700 AV Groningen\label{groningen}
          }
 
   \date{}

 
  \abstract
    {
The assembly history of the stellar component of a massive elliptical galaxy is closely related to that of its dark matter halo.
Measuring how the properties of galaxies correlate with their halo mass can therefore help to understand their evolution.
}
   {
We investigate how the dark matter halo mass of elliptical galaxies varies as a function of their properties, using weak gravitational lensing observations.
To minimise the chances of biases, we focus on the following galaxy properties that can be determined robustly: the surface brightness profile and the colour.
} 
   {
We selected $2409$ central massive elliptical galaxies ($\log{\mstar/M_\odot} \gtrsim 11.4$) from the Sloan Digital Sky Survey spectroscopic sample. We first measured their surface brightness profile and colours by fitting S\'{e}rsic models to photometric data from the Kilo-Degree Survey (KiDS). 
We fitted their halo mass distribution as a function of redshift, rest-frame $r-$band luminosity, half-light radius, and rest-frame $u-g$ colour, using KiDS weak lensing measurements and a Bayesian hierarchical approach.
For the sake of robustness with respect to assumptions on the large-radii behaviour of the surface brightness, we repeated the analysis replacing the total luminosity and half-light radius with the luminosity within a $10$~kpc aperture, $\lten$, and the light-weighted surface brightness slope, $\slope$.
}
   {
We did not detect any correlation between the halo mass and either the half-light radius or colour at fixed redshift and luminosity.
Using the robust surface brightness parameterisation, we found that the halo mass correlates weakly with $\lten$ and anti-correlates with $\slope$.
At fixed redshift, $\lten$ and $\slope$, the difference in the average halo mass between galaxies at the 84th percentile and 16th percentile of the colour distribution is $0.00\pm0.11$~dex.
}
   {
Our results indicate that the average star formation efficiency of massive elliptical galaxies has little dependence on their final size or colour.
This suggests that the origin of the diversity in the size and colour distribution of these objects lies with properties other than the halo mass.
}
   \keywords{Galaxies: elliptical and lenticular, cD --
             Gravitational lensing: weak --
             Galaxies: fundamental parameters
               }

   \maketitle
%

\section{Introduction}\label{sect:intro}

Massive elliptical galaxies, with their smooth stellar distribution and limited presence of cold gas and dust, are relatively simple objects compared to their star-forming counterparts.
Qualitatively speaking, the history of these objects consists of two phases: an initial period of intense star formation, followed by a later stage in which most of their growth is due to mergers with other galaxies \citep{Dok++10, Ose++10}.
However, various aspects of their evolution are still unknown at a quantitative level.

One of the biggest open questions in cosmology is what stopped massive galaxies from forming stars and what prevents them from resuming it.
Feedback from the central supermassive black hole is believed to play an important role in galaxy quenching \citep[i.e. the shutting off of star formation: see][ and references therein]{Fab++12}; however, despite recent progress \citep{Che++16, Bar++18}, demonstrating a causal connection between the two phenomena has proven to be difficult \citep[see for example][]{Sha++20}.

Their post-quenching evolution is also a subject of debate: at fixed stellar mass, elliptical galaxies at $z=0$ are a factor of a few larger, in terms of half-light radius, than their $z=2$ analogues \citep{Dad++05, Tru++07, Bui++08}. Such a size growth is generally attributed to dissipationless mergers \citep{NJO09}, but both the observed and theoretically predicted merger rates appear too low to account for this phenomenon in its entirety \citep{New++12, Nip++12}. 
The addition of new systems to the quiescent galaxy population due to the quenching of previously star-forming galaxies can alleviate this problem if these recently quenched galaxies have larger sizes than the pre-existing ones \citep[see e.g.][]{Sca++07,Dam++19}.
Moreover, the presence of radial gradients in stellar population properties, which were not taken into account in several of the studies mentioned above, act in ways that lead to an overestimate of the size growth \citep{Sue++19a, Sue++19b}.
Nevertheless, we are still lacking a physical model able to reproduce the evolution of the population of massive galaxies in all its complexity.

A fundamental ingredient in the evolution of galaxies is their dark matter halo.
The dark matter distribution determines the gas accretion rate and temperature in the early stages of star formation, as well as the merger rate with other dark matter halos and their respective galaxies.
In particular, the mass of a dark matter halo and its evolution is the most important quantity that determines the properties of the galaxy at its centre \citep{MNW18,Beh++19}.
From a galaxy evolution perspective, having access to measurements of halo mass would then allow one to link the properties of galaxies with their assembly history.
Direct constraints on the dark matter distribution can be used to build more accurate models of the relation between galaxies and halos, which in turn can reduce systematic uncertainties in many cosmological probes that rely on galaxies as tracers of the gravitational potential.
We refer to \citet{W+T18} for a more thorough discussion regarding this point.

One consequence of the connection between the distributions of dark and baryonic matter is the fact that galaxies with higher stellar mass are preferentially found at the centres of the more massive dark matter halos.
This is because the stellar mass of a galaxy is related to the mass of the initial gas reservoir, which is in turn proportional to the dark matter mass.
The exact mapping between stellar and dark matter halo mass, however, depends on the efficiency of the process of conversion of gas into stars. If this quantity varies as a function of the assembly history of a galaxy, additional non-trivial correlations between the properties of galaxies and halo mass could be introduced.
For example, if the gas-to-star conversion efficiency varies with time, the ratio between the stellar and dark matter mass of a galaxy is different among galaxies that formed their stars at different epochs \citep[see for example][]{MNW20}.

Observationally, searches for correlations between halo mass and galaxy properties have been carried out on various fronts.
There have been several efforts aimed at determining how the average halo mass of star-forming galaxies differs from that of quiescent ones at a fixed stellar mass.
These include studies based on weak gravitational lensing \citep{Hoe++05, Vel++14, Hud++15, Man++16, Bil++21, Wan++21}, satellite kinematics \citep{Mor++11, Lan++19}, gas and globular cluster kinematics \citep{P+F21}, galaxy clustering \citep{Zeh++11,Cow++19}, and combinations of multiple such probes \citep{Tin++13}.
While the amplitude and significance of the signal varies among different studies, some of them clearly indicate that passive galaxies tend to be found at the centres of more massive dark matter halos, compared to star-forming galaxies of the same stellar mass.
This suggests that halo mass, or more generally halo assembly history, plays a role in determining the quenching time of a galaxy \citep[see][]{Z+M16}.

\citet{Cha++17}, \citet{SWB19}, and \citet{Hua++20} have attempted to measure the dependence of halo mass on the stellar profile of massive quiescent galaxies at fixed stellar mass, using weak gravitational lensing observations.
They found somewhat conflicting results: \citet{Cha++17} claim a positive correlation between the halo mass and galaxy half-light radius at a fixed stellar mass, but their measurement is in tension with a study by \citet{SWB19}, which ruled out correlations as strong as those reported by \citet{Cha++17}. 
Using data, in large part overlapping with the \citet{SWB19} study, \citet{Hua++20} found a correlation between the halo mass and the ratio between the stellar mass enclosed within two different apertures. They interpreted their measurement as a positive trend between galaxy size and halo mass, qualitatively in agreement with the \citet{Cha++17} work and in disagreement with the \citet{SWB19} study. The origin of this discrepancy is unclear: more robust (model-independent) measurements are needed to settle this issue.

Finally, \citet{Tay++20} used weak lensing data to measure the average dark matter halo mass of galaxies in a narrow stellar mass range around $\log{\mstar/M_\odot} \sim 10.5$ as a function of a set of properties, including the stellar density profile and the star formation rate. 
They found that the halo mass correlates most strongly with the S\'{e}rsic index (positive correlation) and with the half-light radius (negative correlation). 
Their sample, however, included both star-forming galaxies and quiescent ones, so it cannot be compared to the studies cited above directly.

All of these measurements can, in principle, be used to test predictions from theoretical models and hydrodynamical simulations \citep{Sha++14,C+S20,Cui++21}.
Moreover, detecting correlations between the inferred halo mass and galaxy properties at a fixed stellar mass would provide further evidence for the existence of dark matter: if stars were the only mass component, it would be very difficult to explain a variation in the strength of the gravitational lensing signal among galaxies of the same baryonic mass \citep{Bro++21}.
In practice, however, systematic uncertainties in the observations limit our ability to make precise quantitative statements.

A very important source of systematic uncertainty is the measurement of the stellar masses of galaxies.
The standard way to obtain stellar masses is by fitting stellar population synthesis models to the observed photometric or spectroscopic data. Choices in various aspects of these models, such as the duration of each stellar evolutionary stage, the stellar initial mass function (IMF), the dust attenuation, the metallicity and its evolution (or lack thereof), and the star formation history, can introduce biases as large as $0.3$~dex \citep{Con13}.

This systematic uncertainty limits not only the accuracy of our knowledge of the stellar-to-halo mass ratio, but it can also introduce spurious correlations with secondary parameters or, vice versa, erase the signature of correlations.
One illustrative example is the study by \citet{C+S20} of the stellar-to-halo mass relation of Sloan Digital Sky Survey \citep[SDSS][]{Yor++00} galaxies as a function of galaxy morphology.
Using group richness as a proxy for halo mass and taking stellar masses from the catalogue of \citet{Kau++03}, they found that, at a fixed halo mass, disk galaxies have a higher stellar mass compared to ellipticals. However, the signal disappeared when using the stellar mass catalogue of \citet{Cha++15}. Since there is no strong reason to favour one catalogue over the other, it is difficult to draw firm conclusions on the halo mass-galaxy morphology relation with those data.

The goal of this work is to measure the distribution in the dark matter halo mass of central massive elliptical galaxies as a function of observationally robust properties, using weak lensing data from the Kilo Degree Survey \citep[KiDS][]{deJ++13}.
We choose to work with massive galaxies because they are the objects for which the lensing signal is strongest, and we limit the analysis to ellipticals because the regularity of their surface brightness profile allows for more accurate measurements of their structural properties, compared to spiral galaxies.
Finally, we use central galaxies in order to have a cleaner interpretation of the lensing signal. This also implies that our measurements are mostly probing the effects of physical processes internal to the halos, as opposed to the environmental processes to which satellite galaxies are subject (such as stripping or galaxy harassment).

We first focus on the distribution of halo mass as a function of the following quantities: redshift, luminosity, colour, and surface brightness profile.
While the use of luminosity instead of stellar mass can be seen as a step back compared to recent studies on the relation between galaxies and halos, galaxy luminosities do not suffer from the numerous sources of systematic uncertainties mentioned above and can be determined much more accurately than stellar masses.
On the one hand, this approach shifts the burden of predicting the stellar population properties to the theory side, when comparing our measurements to theoretical models. On the other hand, it makes these comparisons more self-consistent because many of the currently available models of galaxy evolution already make predictions on key stellar population properties, such as the star formation history \citep[and hydrodynamical simulations can predict full spectral energy distributions, see for example][]{Tra++17}.

In a second stage, we tackle an additional source of systematic errors: the parameterisation of the stellar surface brightness profile. The total luminosity of a galaxy at a cosmological distance is not directly observable because its surface brightness usually extends out to large radii \citep{T+v11} and falls below the detection limit of current photometric survey data. The total flux is then usually obtained by fitting a model surface brightness profile to the observed data and extrapolating the model to infinity or to a very large radius. This procedure can introduce biases on the derived luminosity and half-light radius, in case the true surface brightness deviates from the model one. \citet{Son20} found that, for a typical massive elliptical galaxy in the SDSS spectroscopic sample imaged at a depth of $\sim26$~mag, 20\% or more of the total flux obtained by fitting a S\'{e}rsic profile is due to extrapolation.

We eliminate this source of systematic errors by employing a more robust description of the surface brightness profile, introduced by \citet{Son20}: instead of the total luminosity and half-light radius, we use the flux within a circularised aperture of radius $10$~kpc and the light-weighted projected stellar density slope in the same aperture.
We then measure the distribution in halo mass as a function of these two quantities, as well as redshift and colour.
These measurements will provide a robust benchmark against which predictions from theoretical models can be tested.

The structure of this paper is as follows. In \Sref{sect:data} we describe the data used for our study, including sample selection criteria. In \Sref{sect:indep} we determine the distribution in luminosity, half-light radius, and colour of our sample. In \Sref{sect:halo} we measure the distribution in dark matter halo mass as a function of redshift, luminosity, half-light radius, and colour. In \Sref{sect:robust} we repeat the analysis by replacing the total luminosity and half-light radius with the aperture luminosity within 10~kpc and the light-weighted surface brightness slope. We discuss our results in \Sref{sect:discuss} and provide our conclusions in \Sref{sect:concl}.

We assume a flat $\Lambda$ cold dark matter cosmology with $\Omega_M=0.3$ and $H_0=70\,\rm{km}\,\rm{s}^{-1}\,\rm{Mpc}^{-1}$. Magnitudes are in AB units, and masses and luminosities are in solar units. Quoted uncertainties correspond to the 68\% credible region.


\section{Data}\label{sect:data}

\subsection{Sample selection}\label{ssec:sample}

Our goal is to select a sample consisting of quiescent galaxies with an elliptical morphology and available spectroscopic data, with which to obtain precise redshift, luminosity, rest-frame colour, and size measurements. 
We also wish to carry out a weak lensing analysis of the same sample. For this purpose, we require our galaxies to lie in regions of the sky covered by the Data Release 4 (DR4) of KiDS \citep{Kui++19}. 
For the sake of a clean interpretation of our results, particularly from the weak lensing part, we wish to limit our study to central galaxies. 
We achieved these goals in a series of steps, as described below.

We selected galaxies from the spectroscopic sample of the SDSS Data Release 12 \citep[DR12,][]{Ala++15}. In particular, we considered objects with spectra taken with the SDSS spectrograph \citep{Sme++13} belonging to the Luminous Red Galaxy sample \citep{Eis++01} of the SDSS Legacy Survey.
We first applied a minimum redshift condition, selecting only objects at $z>0.15$. This cut is motivated by the fact that the galaxy stellar mass function of the SDSS Luminous Red Galaxy sample is more or less independent of redshift above $z=0.15$, by construction, while it is a strong function of redshift below this limit. This makes it easier to separate trends with a redshift from trends with luminosity.

To select quiescent galaxies, we applied a cut to the H$\alpha$ equivalent width $\ewha$, selecting only objects with $\ewha > -3\AA$ (the negative equivalent width corresponds to emission). This step, together with our minimum redshift condition, resulted in a sample of $\sim3700$ luminous red galaxies at $z>0.15$ with KiDS DR4 photometry.
We removed obvious disk-like objects by eliminating galaxies with a projected $r-$band minor-to-major axis ratio smaller than $0.5$, as measured from KiDS photometric data (more details on these measurements are provided in Section \ref{ssec:sb}): they account for 8\% of the sample.

We then removed $\sim200$ objects that are identified as cluster members with a probability being a central galaxy smaller than $0.5$ in the red-sequence Matched-filter Probabilistic Percolation (redMaPPER) catalogue \citep{Ryk++14}.
Using KiDS colour-composite images in $gri$ bands, we visually inspected the remaining objects, and we eliminated galaxies with spiral arms, with disk or ring features, or those in very close proximity to objects with a comparable or higher brightness. This visual inspection step removed a further 15\% of the galaxies in the sample.
Finally, we applied a galaxy isolation cut, which was needed in order to render the weak lensing analysis computationally tractable. For each galaxy, we drew a cone around it with a 500~kpc physical radius at its redshift. From the sample, we then removed any galaxy where the cone overlapped with that of another more luminous galaxy. 
In \Sref{sect:halo} we give more details on the rationale for this choice.
Our final sample consists of $2409$ galaxies spanning the redshift range $0.15 < z < 0.4$.
The tenth percentile, median, and 90th percentile of the distribution in $\log{\mstar}$ of the sample (measured as described in Section \ref{ssec:luminosity}) are $11.47$, $11.68$, and $11.93$, respectively.

\subsection{KiDS photometry}

Our measurements of the surface brightness profiles of our galaxies and the weak lensing data are based on photometric observations from KiDS DR4. KiDS is a photometric survey carried out from the ESO VLT Survey Telescope \citep[VST][]{C+S11,Cap++12}, with the main goal of obtaining cosmological constraints from cosmic shear measurements. KiDS DR4 covers an area of approximately 1,000 square degrees, imaged in $u, g, r,$ and $ i$ bands with the instrument OmegaCAM \citep{Kui++11}. In addition to optical data, KiDS DR4 includes matched-photometry measurements in the near-infrared bands $Z, Y, J, H$, and $K_s$ obtained from the VISTA telescope as part of the VISTA Kilo-degree INfrared Galaxy survey \citep[VIKING][]{Edg++13}.
The $r-$band data is the one with the best image quality, with a typical seeing $\mathrm{FWHM} < 0.8''$, while the 5$\sigma$ mean limiting magnitude within a $2''$ aperture is $\approx25$.
For this reason, we used the $r-$band to measure the structural parameters of our galaxies, as described in the next section.
We refer readers to \citet{Kui++19} for details on the KiDS DR4 data reduction.

\subsection{Surface brightness profile measurements}\label{ssec:sb}

We describe the surface brightness distribution of each galaxy with an elliptical S\'{e}rsic profile,
\begin{equation}\label{eq:sersic}
I(x,y) = I_0\exp{\left\{-b(n)\left(\frac{R}{\reff}\right)^{1/n}\right\}},
\end{equation}
where $x$ and $y$ are Cartesian coordinates with an origin at the galaxy centre and aligned with the galaxy major and minor axes, respectively, and $R$ is the circularised radius
\begin{equation}\label{eq:circularised}
R^2 \equiv qx^2 + \frac{y^2}{q},
\end{equation}
where $q$ is the axis ratio, $n$ is the S\'{e}rsic index, and $b(n)$ is a numerical constant defined in such a way that the isophote of radius $R=\reff$ encloses half of the total light \citep[see][]{C+B99}.
Throughout this paper, we refer to $\reff$ as the half-light radius and as the size of a galaxy, interchangeably.
The best-fit S\'{e}rsic parameters for each galaxy were obtained by fitting seeing-convolved S\'{e}rsic models to sky-subtracted and coadded $r-$band images using the software {\sc 2dphot} \citep{LaB++08}, largely following the procedure adopted by \citet{Roy++18} for the analysis of data from previous KiDS data releases. 

We obtained magnitudes in all nine bands by considering Galactic extinction-corrected colours, measured with the Gaussian Aperture and point spread function (PSF) technique \citep[GAaP][]{Kui++15}, and using them to rescale the S\'{e}rsic fit-based $r$-band flux accordingly. An implicit assumption in this step is that colours do not vary across the spatial extent of a galaxy.
We discuss the possible impact of this assumption in Section \ref{ssec:systematics}.

\subsection{Rest-frame colour and luminosity measurements}\label{ssec:luminosity}

We base our analysis on measurements of the rest-frame $r$-band luminosity and $u-g$ colour.
We obtained these quantities by fitting spectral energy distribution templates, obtained with the software {\sc Galaxev} \citep{BC03}, to the observed KiDS multi-band data.
In particular, we generated composite stellar population synthesis models assuming a Chabrier stellar IMF \citep{Cha03} and an exponentially declining star formation history, a range of values of metallicity, and dust attenuation. 
We then optimised for the parameters of the model given the observed 9-band magnitudes to obtain a best-fit spectral template, with which we derived the rest-frame $r$-band luminosity and $u-g$ colour.

The model is not able to fit the observed magnitudes down to the noise level, which is very low (typically a few millimags in all bands, except $u$).
The derived $u-g$ colour, then, is sensitive to the spectral energy distribution of the galaxy over the full wavelength range probed by KiDS photometry, and it should therefore be considered as an average measure of galaxy redness over a broad spectral range.
To make sure that the results are independent of our way of measuring $u-g$, we repeated the analysis using a more direct estimator of rest-frame colour, the $D_n4000$ spectral index, which we obtained from the SDSS spectroscopic catalogue. The $D_n4000$ index is defined as the ratio between the continuum level redward and blueward of the $4000$~$\AA$ break \citep[see][ for details]{Bal++99}.
Our measurements of the distribution of halo mass as a function of $D_n4000$ are indistinguishable from those obtained using $u-g$. Therefore, we only show the latter in this paper.

One by-product of this spectral energy distribution fitting process is an estimate of the stellar mass of each galaxy. 
We chose not to use this information for the core of our analysis since the stellar mass parameter is highly degenerate with the stellar IMF, the metallicity, and the star formation history parameters, and because the way that these aspects of our composite stellar population models are set is somewhat arbitrary.
The only aspect of our study that relies on these stellar mass measurements is the estimate of the (small) baryonic contribution to the weak lensing signal, in \Sref{sect:halo}.

\subsection{Aperture luminosity and light-weighted slope}

We also consider an alternative parameterisation of the luminosity profile of our galaxies, based on quantities that are measured more robustly than the total luminosity and the half-light radius.
We use the rest-frame $r$-band luminosity enclosed within a circularised aperture of physical size $10$~kpc, $\lten$, and the light-weighted surface brightness slope within the same aperture, $\slope$. Following \citet{Son20}, the latter is defined as
\begin{equation}\label{eq:slopedef}
\slope \equiv - \dfrac{2\pi\int_0^{10}R\dfrac{d\log I_r}{d\log R}I_r(R)dR}{2\pi\int_0^{10}RI_r(R)dR} = 2 - \dfrac{2\pi(10)^2I_r(10)}{\lten},
\end{equation}
where $I_r(R)$ is the circularised rest-frame $r$-band surface brightness profile of the galaxy.

To calculate $\lten$ and $\slope$, we assume that the true surface brightness profile of each galaxy is given by its best-fit S\'{e}rsic model. This is a good approximation as the typical deviation in the surface brightness profile of massive elliptical galaxies from a S\'{e}rsic model, in the region constrained by the data, is a few percent \citep[see Figure 4 of][]{Son20}.
Although based on the same S\'{e}rsic model used to estimate $L_r$ and $\reff$, the advantage of using $\lten$ and $\slope$ is that they do not rely on any extrapolation to large radii where the surface brightness profile is unconstrained, and they are therefore determined more robustly.

\subsection{Background source shape and redshift measurements}

Our weak lensing analysis is based on shape measurements of galaxies in the background of the objects in our sample, obtained by \citet{Gib++21}.
Shapes are described by two ellipticity components, $(\epsilon_1, \epsilon_2)$, which were measured from individual $r-$band exposures with the {\sc lensfit} algorithm \citep{Mil++07,Mil++13,Fen++17}. 
{\sc lensfit} also assigns a weight $w_s$ to each source, which takes into account the uncertainty in the ellipticity measurements and the intrinsic scatter in the ellipticity distribution of the source sample. The latter dominates the error budget for most sources \citep[see][ for details]{Mil++13}.
As we explain in \Sref{sect:halo}, shape measurements are weighted by their $w_s$ in the weak lensing analysis.
We estimated multiplicative shear biases using image simulations from \citet{Kan++19}. The uncertainties on these are of the order of $10^{-3}$.

Background sources were selected on the basis of their photometric redshifts (photo-zs). These were obtained using the Bayesian Photometric Redshift code \citep[{\sc BPZ}][]{Ben00,Coe++06} on KiDS $u, g, r,$ and $i$ and VIKING $Z, Y, J, H,$ and $ K_s$ PSF-matched photometry, using a source magnitude prior from \citet{Rai++14}. We refer readers to \citet{Wri++19} for details on the photo-z measurements.
For each lens at redshift $\zlens$, we selected only galaxies with a maximum a posteriori redshift $z_\mathrm{B}$ larger than $\zlens+0.2$. 
Following \citet{Hil++21}, we then determined the weighted redshift distribution of the resulting sample, $n_w(\zsource)$, incorporating the {\sc Lensfit} weight. This was done with a set of objects with spectroscopic redshift measurements which was matched to the source sample with a self-organising map approach \citep{Wri++20}. The redshift distribution of the spectroscopic sample was re-weighted to match the relative abundance of source galaxies with different photometric measurements, multiplied by their {\sc lensfit} weight.

\section{Distribution in luminosity, half-light radius, and colour}\label{sect:indep}

\begin{figure*}
\includegraphics[width=\textwidth]{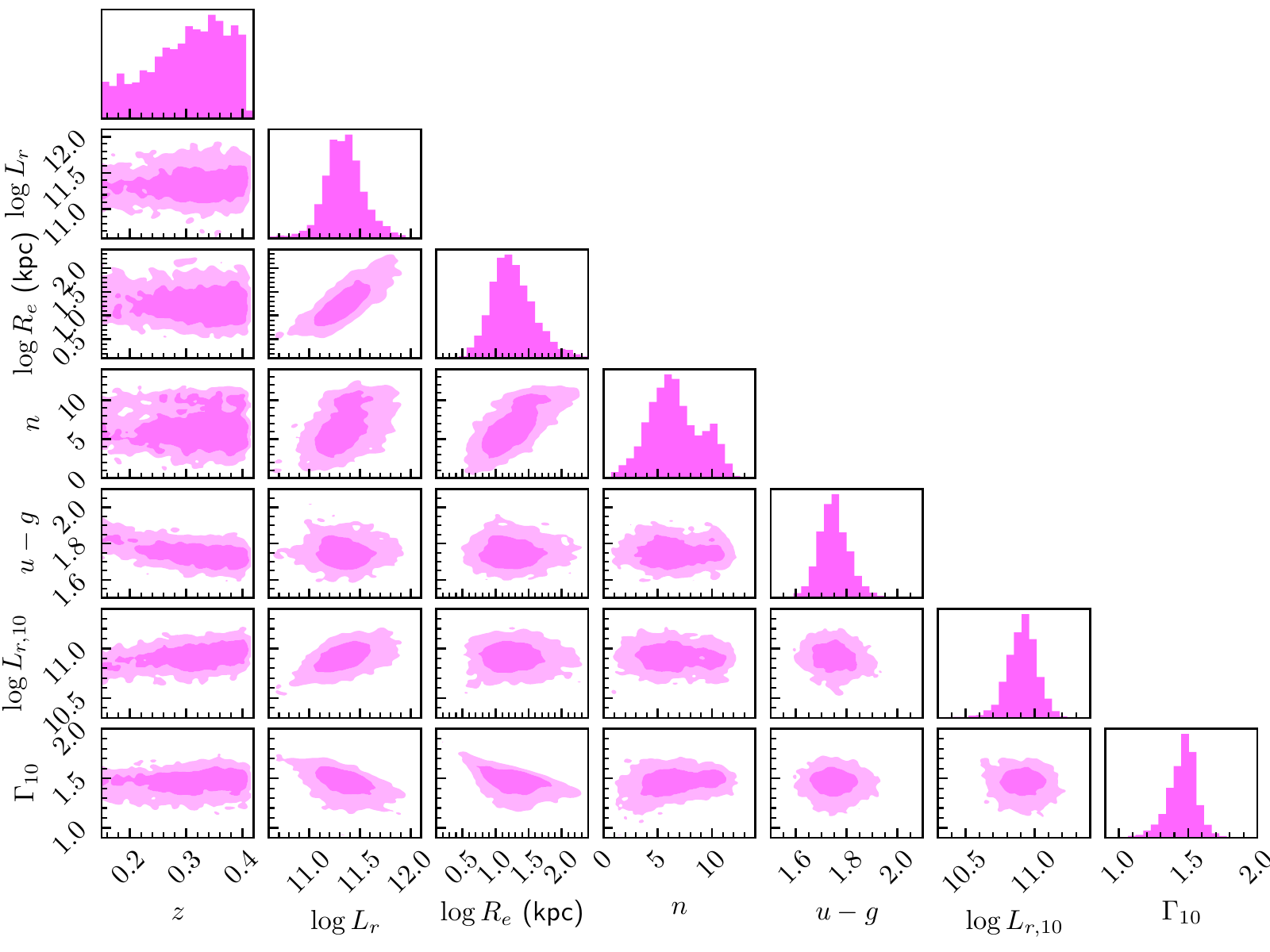}
\caption{
Distribution in redshift, rest-frame $r$-band luminosity, half-light radius, S\'{e}rsic index, rest-frame $u-g$ colour, luminosity within $10$~kpc, and light-weighted surface brightness slope within $10$~kpc of the galaxy sample.
Contour levels enclose 68\% and 95\% of the sample, respectively.
\label{fig:samplecp}
}
\end{figure*}

In \Fref{fig:samplecp} we show the distribution in redshift, luminosity, half-light radius, S\'{e}rsic index, rest-frame $u-g$ colour, $\lten$, and $\slope$ of our galaxy sample.
We can clearly see that $\reff$ correlates positively with $L_r$ and $n$, while $\slope$ anti-correlates with $\lten$ \citep[as measured by][ on a similar sample of elliptical galaxies]{Son20}.
The distribution in the S\'{e}rsic index, however, shows bimodality, with a secondary peak at around $n=10$.
It is not clear whether this bimodality is real, or if it is a spurious result of the fit with {\sc 2dphot}.
Visual inspection did not reveal any obvious difference between galaxies with $n>9$ and the rest of the sample, and we found no correlation between $n$ and the $\chi^2$ of the photometric fit.

Galaxies with a very large S\'{e}rsic index can be problematic because a large fraction of their model flux is due to an extrapolation of the surface brightness profile to regions that are below the detection limit of the survey \citep[see Figure 2 of ][]{Son20}.
However, as we explained in \Sref{sect:intro}, the parameterisation in terms of $\lten$ and $\slope$ is robust with respect to this extrapolation.
For these reasons, we decided not to apply any further cut, and we carry out the analysis with all galaxies, regardless of their S\'{e}rsic index.

As a first step in our analysis, we are interested in compressing these data into a few numbers describing how the size and the colour of our galaxies correlate with each other, how they scale with redshift and luminosity, and what the scatter is around their mean relations.
For this purpose, we fitted a bivariate Gaussian in $\log{\reff}$ and rest-frame $u-g$ colour to the data. 
We chose this model because it is the simplest one allowing us to infer the above quantities. 
We quantify the goodness of the Gaussian assumption later in this section.
Redshift and luminosity are treated as independent variables and their distribution is not modelled.

We write our model as follows (for the sake of keeping the notation compact, we indicate the rest-frame $u-g$ colour as $C$):
\begin{equation}\label{eq:indepmodel}
{\rm P}(\log{\reff},C) = \mathcal{R}(\reff|z,\lumr)\mathcal{K}(C|z,\lumr,\reff).
\end{equation}
In the above equation, the factor $\mathcal{R}$ describes the distribution in half-light radius as a function of redshift and luminosity. 
This is a Gaussian in $\log{\reff}$
\begin{equation}\label{eq:reffmodel}
\mathcal{R}(\log{\reff}|z,\lumr) = \frac{1}{\sqrt{2\pi}\sigmaR}\exp{\left\{-\frac{\left(\log{\reff} - \muR(z,\lumr)\right)^2}{2\sigmaR^2}\right\}},
\end{equation}
with mean
\begin{equation}\label{eq:meansize}
\muR(z,\lumr) = \muRzero + \zetaR(z - 0.3) + \betaR(\log{\lumr} - 11.3)
\end{equation}
and dispersion $\sigmaR$.
In \Eref{eq:meansize}, the parameter $\muRzero$ indicates the average value of $\log{\reff}$ at the pivot point $z=0.3$ and $\log{\lumr}=11.3$, while coefficients $\zetaR$ and $\betaR$ describe a linear scaling of $\log{\reff}$ with $z$ and $\log{\lumr}$, respectively.

The factor $\mathcal{K}$ describes the distribution in colour as a function of redshift, luminosity, and half-light radius. 
This is also a Gaussian distribution
\begin{equation}\label{eq:colourdist}
\mathcal{K}(C|z,\lumr,\log{\reff}) = \frac{1}{\sqrt{2\pi}\sigmaC}\exp{\left\{-\frac{\left(C - \muC(z,\lumr,\reff)\right)^2}{2\sigmaC^2}\right\}},
\end{equation}
with the mean that scales with redshift, luminosity, and half-light radius as
\begin{equation}\label{eq:meancolour}
\begin{split}
\muC(z,\lumr,\reff) = & \muCzero + \zetaC(z - 0.3) + \betaC(\log{\lumr} - 11.3) + \\ & \xiC\left(\log{\reff} - \muR(z,\lumr)\right)
\end{split}
\end{equation}
and dispersion $\sigmaC$.
We would like to point out that in \Eref{eq:meancolour}, the pivot point in $\log{\reff}$ space is not fixed, but it is set to the average half-light radius $\muR(z,\lumr)$, defined in \Eref{eq:meansize}, which scales with redshift and luminosity. This means that $\xiC$ indicates how colour scales with relative size, compared to the average value of $\log{\reff}$ of galaxies with the same redshift and luminosity.

We fitted the parameters of the model with a Markov chain Monte Carlo (MCMC), using the affine-invariant sampling method of \citet{G+W10}, implemented in Python with the package {\sc emcee} \citep{For++13}.
The statistical uncertainties on all of the observed quantities, redshift, luminosity, size, and colour are very small (the typical photometric error is $0.01$~mag); therefore, we neglected them when carrying out the fit.
After performing the fit, we removed the outliers, which we identified with a sigma-clipping procedure, and we repeated the fit on the new sample, iterating until convergence.
To be precise, we labelled as outliers any objects with values of $\log{\reff}$ that are more than $4\sigmaR$ away from the average size $\muR(z,\lumr)$ or with values of $C$ that are more than $4\sigmaC$ away from the average colour $\muC(z,\lumr,\reff)$. We removed 28 galaxies from the sample with this procedure. 
We visually inspected these galaxies and found that, for the most part, they are objects with a bright object in their proximity: we suspect that either the GAaP or S\'{e}rsic model fits, on which our photometry is based, suffered from a catastrophic failure due to blending.

The median value and 68\% credible limits on the marginal posterior probability distribution of each model parameter are reported in \Tref{tab:indep}.
The results can be summarised as follows. Size scales with luminosity with a steeper-than-linear relation, $\betaR=1.381\pm0.018$, and it decreases with increasing redshift at a fixed luminosity: $\zetaR=-0.79\pm0.05$. This is a steeper evolution than the change in size at a fixed stellar mass, as reported in the literature: for example, \citet{New++12} observed an average size evolution rate at fixed $\mstar$ of $d\left<\log{\reff}\right>/dz\approx-0.12$ over the redshift range $0 < z < 1$. The difference is due to the luminosity evolution of galaxies at a fixed stellar mass:  
at a fixed luminosity, passive galaxies at a higher redshift are intrinsically less massive than their lower redshift counterparts, and therefore their sizes are smaller on average. 
Our measurement of the intrinsic scatter in $\log{\reff}$ at a fixed redshift and luminosity is $\sigmaR=0.183\pm0.003$.

The rest-frame $u-g$ colour decreases with an increasing redshift at a fixed luminosity, $\zetaC=-0.411\pm0.016$, which is qualitatively in agreement with a passive evolution scenario, and almost independent of luminosity at a fixed redshift, $\betaC=-0.008\pm0.005$.
Moreover, at a fixed redshift and luminosity, the colour is independent of excess size: $\xiC=-0.008\pm0.006$.

\begin{table*}
\caption{Parameters describing the size and colour distribution as a function of redshift and luminosity.
}
\label{tab:indep}
\begin{tabular}{cccl}
\hline
\hline
Parameter & Prior & Med.$\pm1\sigma$ & Description \\
\hline
$\muRzero$ & $U(0,2)$ & $1.212\pm0.004$ & Mean $\log{\reff}$ at $z=0.3$ and $\log{\lumr}=11.3$. \\
$\zetaR$ & $U(-3,3)$ & $-0.80\pm0.05$ & Linear scaling of the average $\log{\reff}$ with redshift. \\
$\betaR$ & $U(-3,3)$ & $1.374\pm0.017$ & Linear scaling of the average $\log{\reff}$ with $\log{\lumr}$. \\
$\sigmaR$ & $U(0,1)$ & $0.183\pm0.003$ & Intrinsic scatter in $\log{\reff}$ at fixed redshift and luminosity. \\
$\muCzero$ & $U(1,3)$ & $1.751\pm0.001$ & Mean $u-g$ colour at $z=0.3$, $\log{\lumr}=11.3$ and average size. \\
$\zetaC$ & $U(-10,10)$ & $-0.412\pm0.015$ & Linear scaling of the average colour with redshift. \\
$\betaC$ & $U(-3,3)$ & $-0.009\pm0.005$ & Linear scaling of the average colour with $\log{\lumr}$. \\
$\xiC$ & $U(-10,10)$ & $-0.007\pm0.006$ & Linear scaling of the average colour with excess size. \\
$\sigmaC$ & $U(0,1)$ & $0.053\pm0.001$ & Intrinsic scatter in colour at fixed redshift, luminosity and size. 
\end{tabular}
\tablefoot{
The parameters are introduced in \Eref{eq:reffmodel}, \ref{eq:meansize}, \ref{eq:colourdist}, and \ref{eq:meancolour}.
Column (2): priors on the parameters.
Column (3): median and 68\% credible region of the marginal posterior probability distribution of each parameter given the data.
}
\end{table*}

This finding is somewhat surprising, given previous similar studies from the literature. For instance, \citet{Kim++18} found a positive, albeit weak, correlation between the stellar surface mass density of a sample of spheroidal galaxies and the $D_n4000$ spectral index. 
\citet{Z+G17} also report a correlation between $D_n4000$ and the size at a fixed stellar mass in a sample of quiescent galaxies.
In order to compare our finding with these studies, we repeated our analysis by using $D_n4000$ measurements in place of the colour. We found that $D_n4000$ has a weak positive correlation with luminosity, thus still in apparent discrepancy with the studies mentioned above.
As we show in Section \ref{ssec:mltrend}, however, using stellar mass in place of luminosity leads to an anti-correlation between size and colour.
Moreover, the apparent discrepancy with respect to the \citet{Z+G17} and \citet{Kim++18} studies could be due to differences in the sample selection criteria: for example, the \citet{Z+G17} sample is dominated by galaxies with a stellar mass below $10^{11}M_\odot$, which are excluded by our luminosity cut.

\begin{figure*}
\begin{tabular}{cc}
\includegraphics[width=\columnwidth]{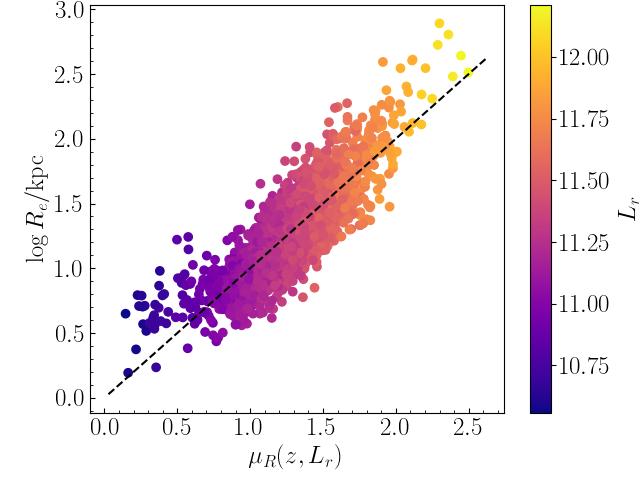}
&
\includegraphics[width=\columnwidth]{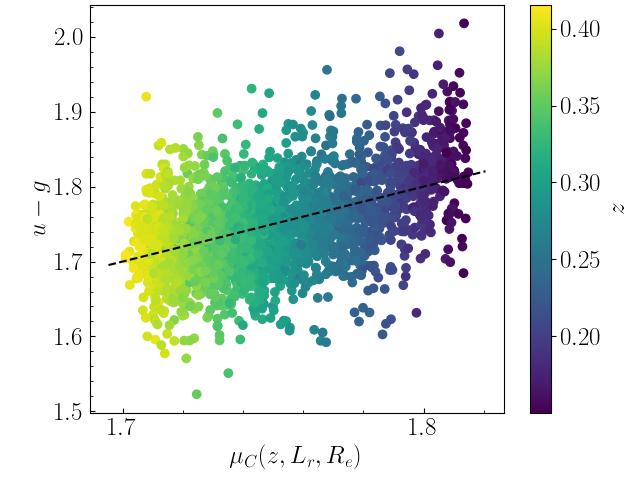}
\end{tabular}
\caption{Comparison between observed and model-predicted sizes and colours.
Left: Observed half-light radius as a function of the average $\reff$ of galaxies with the same redshift and luminosity, as predicted by \Eref{eq:meansize}.  Right: Observed rest-frame $u-g$ colour as a function of the average colour of galaxies with the same redshift, luminosity, and size, as predicted by \Eref{eq:meancolour}. Uncertainties on the model parameters are neglected, as they are very small with respect to the observational scatter.
The dashed line marks the $1:1$ relation.
\label{fig:indep}
}
\end{figure*}

The measurements shown in this section were obtained under the assumption of a bi-variate Gaussian distribution in $\log{\reff}$ and $C$ at a fixed redshift and luminosity.
In order to verify the accuracy of this model, in \Fref{fig:indep} we show the observed size and colour distributions as a function of the predicted average value of these quantities, obtained via \Eref{eq:meansize} and \Eref{eq:meancolour}.
The model is able to correctly reproduce the average size and colour of the majority of the sample. 
However, there is a subset of galaxies at low luminosity, redder colours, and low redshift that appear to deviate from the $1:1$ line. We repeated the analysis by removing objects with $z<0.2$. While this additional selection caused the inferred distribution in size and colour to change, the final measurements of the halo mass distribution did not vary. Therefore, we present the results obtained with the full sample.

To check the robustness of the Gaussian scatter assumption, we measured the fraction of galaxies with values of $\reff$ that are within $1\sigmaR$ of $\mu_R(z,L_r)$. This is 71\%, which is slightly larger than the expected value of 68\% of a pure Gaussian. Similarly, the fraction of galaxies with colour within $1\sigmaC$ of $\mu_C(z,L_r,\reff)$ is 71\%. These small deviations from a Gaussian distribution are not concerning for the goals of this work.

\section{Dark matter halo mass distribution}\label{sect:halo}

\subsection{Model}

We wanted to measure the distribution in dark matter halo mass as a function of redshift, luminosity, size, and colour. We did so with a Bayesian hierarchical inference approach, following the method of \citet{S+L18}.
In order to fit the weak lensing data, we needed to model the projected surface mass density distribution of the lens sample.
We only used data covering relatively small scales, $50-500$~kpc, where the lensing signal from the lenses themselves dominates over the contribution from the large-scale structure. 
Therefore, we modelled each lens simply as the sum of a dark matter halo and a stellar component.
We describe the former with a spherical Navarro, Frenk, \& White profile \citep[NFW,][]{NFW97}, and the latter with a circularised S\'{e}rsic profile.
We ignored any contribution from gas.
We discuss the impact of these assumptions in Section \ref{ssec:systematics}.

The 3D density distribution of a spherical NFW profile, as a function of radius $r$, is the following:
\begin{equation}
\rho_{\mathrm{DM}}(r) \propto \frac{1}{r(1+r/r_\mathrm{s})^2}.
\end{equation}
It has two degrees of freedom, related to the normalisation and the scale radius $r_\mathrm{s}$.
We describe the normalisation in terms of the mass $\mhalo$ enclosed within the spherical shell of radius $\rhalo$, within which the average density is $200$ times the critical density of the Universe at the redshift of the galaxy, $\rho_\mathrm{c}$. 
We refer to $\mhalo$ simply as the halo mass.
As a second parameter, we used the concentration $\chalo$, defined as the ratio between $\rhalo$ and the scale radius.

We then modelled the distribution in $\mhalo$ and $\chalo$ of our lens population with the following functional form:
\begin{equation}\label{eq:dmdist}
\begin{split}
\pr(\log{\mhalo},\log{\chalo}|z,\lumr,\reff,C) = & \mathcal{H}(\log{\mhalo}|z,\lumr,\reff,C)\times \\
& \mathcal{C}(\log{\chalo}|\mhalo).
\end{split}
\end{equation}
We modelled the first term as a Gaussian in $\log{\mhalo}$,
\begin{equation}\label{eq:mhalodist}
\mathcal{H}(\log{\mhalo}|z,\lumr,\reff,C) = \frac{1}{\sqrt{2\pi}\sigma_h}\exp{\left\{-\frac{\left(\log{\mhalo} - \mu_h\right)^2}{2\sigma_h^2}\right\}},
\end{equation}
with mean
\begin{equation}\label{eq:meanhalo}
\begin{split}
\mu_h(z,\lumr,\reff,C) = & \mu_{h,0} + \zetah(z - 0.3) + \betah(\log{\lumr} - 11.3) + \\
& \xih\left(\log{\reff} - \mu_R(z,\lumr)\right) + \\ 
& \nuh\left(C - \mu_C(z,\lumr,\reff)\right)
\end{split}
\end{equation}
and dispersion $\sigma_h$.
Similarly to our parameterisation of the colour distribution in \Sref{sect:indep}, in \Eref{eq:meanhalo} we allowed for the average $\log{\mhalo}$ to scale with redshift, luminosity, and excess size, via the parameters $\zetah$, $\betah$, and $\xih$. In addition, we allowed for a correlation between the halo mass and excess colour $C - \mu_C(z,\lumr,\reff)$, which is the difference between the colour of a galaxy and the average colour of galaxies of the same redshift, luminosity, and size, via parameter $\nuh$.
\citet{S+L18} show how a log-Gaussian distribution in $\mhalo$ of the same kind as that found in \Eref{eq:mhalodist} is able to accurately capture the average halo mass of a sample of massive galaxies and the correlations between $\mhalo$ and galaxy properties, when fitted to weak lensing data similar to those of the KiDS survey.

Then, we described the term $\mathcal{C}$ in \Eref{eq:dmdist} with a fixed mass-concentration relation for the distribution in $\chalo$:
\begin{equation}\label{eq:c200dist}
\mathcal{C}(\chalo|\mhalo) = \dfrac{1}{\sqrt{2\pi}s_c}\exp{\left\{-\dfrac{(\log{\chalo}-\mu_c(\mhalo))^2}{2s_c^2}\right\}},
\end{equation}
with mean
\begin{equation}\label{eq:c200mu}
\mu_\mathrm{c}(\mhalo) = \mu_{\mathrm{c},0} + \beta_{\mathrm{c}}(\log{\mhalo} - 12)
\end{equation}
and dispersion $s_\mathrm{c}$.
Following the N-body simulation predictions of \citet{Mac++08} based on WMAP5 cosmology, we fixed $\mu_{\mathrm{c},0}=0.83$, $\beta_\mathrm{c} = -0.097$, and $s_\mathrm{c}=0.1$.
We discuss the impact of this choice in Section \ref{ssec:systematics}.

Our model distribution in dark matter halo mass is then summarised by the following set of free parameters:
\begin{equation}\label{eq:hyperpars}
\hyperp \equiv \left\{\muhzero, \zetah, \betah, \xih, \nuh, \sigma_h\right\}.
\end{equation}
In principle, $\muR$ and $\muC$ are also free parameters that enter the halo mass distribution through \Eref{eq:meanhalo}. In practice, however, we fixed their values to the median of their marginal posterior distribution obtained from the analysis of \Sref{sect:indep}, for the purpose of reducing the dimensionality of the problem. This is a fair approximation since the uncertainties on $\muR$ and $\muC$ are very small (see \Tref{tab:indep}).

We wanted to determine the posterior probability distribution of the model parameters $\hyperp$ given the data $\data$. Using Bayes' theorem, this is given by
\begin{equation}
\pr(\hyperp|\data) \propto \pr(\hyperp)\pr(\data|\hyperp),
\end{equation}
where $\pr(\hyperp)$ is the prior probability on the parameters and $\pr(\data|\hyperp)$ is the likelihood of observing the data given the model.

Weak lensing data consist of an ensemble of shape measurements on galaxies located behind the lens. The main observable is the complex ellipticity, $\compell = (\ellone,\elltwo)$. The expectation value of the tangential component (with respect to the lens galaxy) of the ellipticity $\compellt$ is the tangential reduced shear at the image position $\boldsymbol\theta$ and source redshift $z_s$, $\redsheart(\boldsymbol\theta,z_s)$, corrected for the shear multiplicative bias $(1+m)$:
\begin{equation}\label{eq:et_exp}
\mathrm{E}\left[\compellt\right] = (1+m)\redsheart(\boldsymbol\theta,\zsource).
\end{equation}

To predict the value of the reduced shear at each source position given our model, we worked under the isolated lens assumption of \citet{S+L18}: we assumed that only the lens galaxy and relative dark matter halo contribute to the lensing signal on sources around it. This allows us to predict the value of the reduced shear $\redsheart$ given the halo mass, the halo concentration, the redshift, and the stellar mass of a lens galaxy.
We then approximated the likelihood of a single shape measurement $\epsobs$ as a Gaussian centred on \Eref{eq:et_exp} and with a dispersion equal to the inverse square root of the {\sc lensfit} weight, $w_s^{-1/2}$.
For a given source redshift $z_s$, the shape measurement likelihood is then
\begin{equation}\label{eq:singlewllike}
\pr(\epsobs|\mhalo,\chalo,z_s) = \mathcal{N}((1+m)\redsheart(\boldsymbol\theta,z_s,\mhalo,\chalo),w_s^{-1}),
\end{equation}
where $\mathcal{N}(\mu,\sigma^2)$ indicates a Gaussian distribution with mean $\mu$ and variance $\sigma^2$.
The exact individual redshifts of the source galaxies, however, are not known. We thus marginalise \Eref{eq:singlewllike} over the {\sc lensfit} weight-weighted source redshift distribution, $n_w(z_s)$:
\begin{equation}
\pr(\epsobs|\mhalo,\chalo) = \int dz_s n_w(z_s) \mathcal{N}((1+m)\redsheart(\boldsymbol\theta,z_s,\mhalo,\chalo),w_s^{-1}),
\end{equation}
where the model reduced shear term in the integrand is zero for values of $z_s$ that are smaller than the lens redshift.

The function $n_w(z_s)$ was determined using a spectroscopic sample of galaxies matched to the entire KiDS DR4 shape measurements catalogue \citep{Hil++21}. 
This is not necessarily the same as the redshift distribution of the sources around our lenses as our source sample could be contaminated by galaxies physically associated with the local environment of the lenses. 
To correct for this difference, we first made five radial bins, then compared the number density of sources around our lens sample with that around a large set of random points covering the same footprint as the lenses. We found that the former is on average 10\% larger than the latter, with a mild dependence on the radius. Using these measurements, we corrected the $n_w(z_s)$ of the sources in each radial bin by boosting the number density of sources (which are not lensed) at the same redshift of the lens.

The full likelihood for an ensemble of shape measurements around a single lens is the product over the sources of terms of the form of \Eref{eq:singlewllike}.
We only considered sources in an annulus of physical radius at the lens redshift between 50 and 500~kpc.
The isolation criterion applied in the construction of the sample, which is described in Section \ref{ssec:sample}, ensures that we never use the same shape measurement more than once in different lenses.
Additionally, it allows us to write the likelihood as the product of $N$ independent factors, with $N$ being the number of lenses in our sample:
\begin{equation}\label{eq:integrals}
\pr(\data|\hyperp) = \prod_i^N\pr(\datai|\hyperp),
\end{equation}
where $\datai$ are the shape measurements around the $i$-th lens.
As explained above, the shear data depend on the values of the halo mass and concentration. To evaluate each term in the product of \Eref{eq:integrals}, it is then necessary to average over all possible values taken by these quantities, as predicted by the following model:
\begin{equation}\label{eq:integral}
\pr(\datai|\hyperp) = \int d\log{\mhalo} d\log{\chalo} \pr(\datai|\mhalo,\chalo) \pr(\mhalo,\chalo|\hyperp),
\end{equation}
where $\pr(\log{\mhalo},\log{\chalo}|\hyperp)$ is the model probability distribution of \Eref{eq:dmdist}.

We assumed flat priors on each parameter, with bounds reported in \Tref{tab:dminfer}. We sampled the posterior probability distribution $\pr(\hyperp|\data)$ with an MCMC. At each draw of the set of parameters $\hyperp$, we evaluated the integrals of \Eref{eq:integral} with Monte Carlo integration and importance sampling, following \citet{Son++19b} (see their Appendix A).

\subsection{Results}

\begin{figure*}
\includegraphics[width=\textwidth]{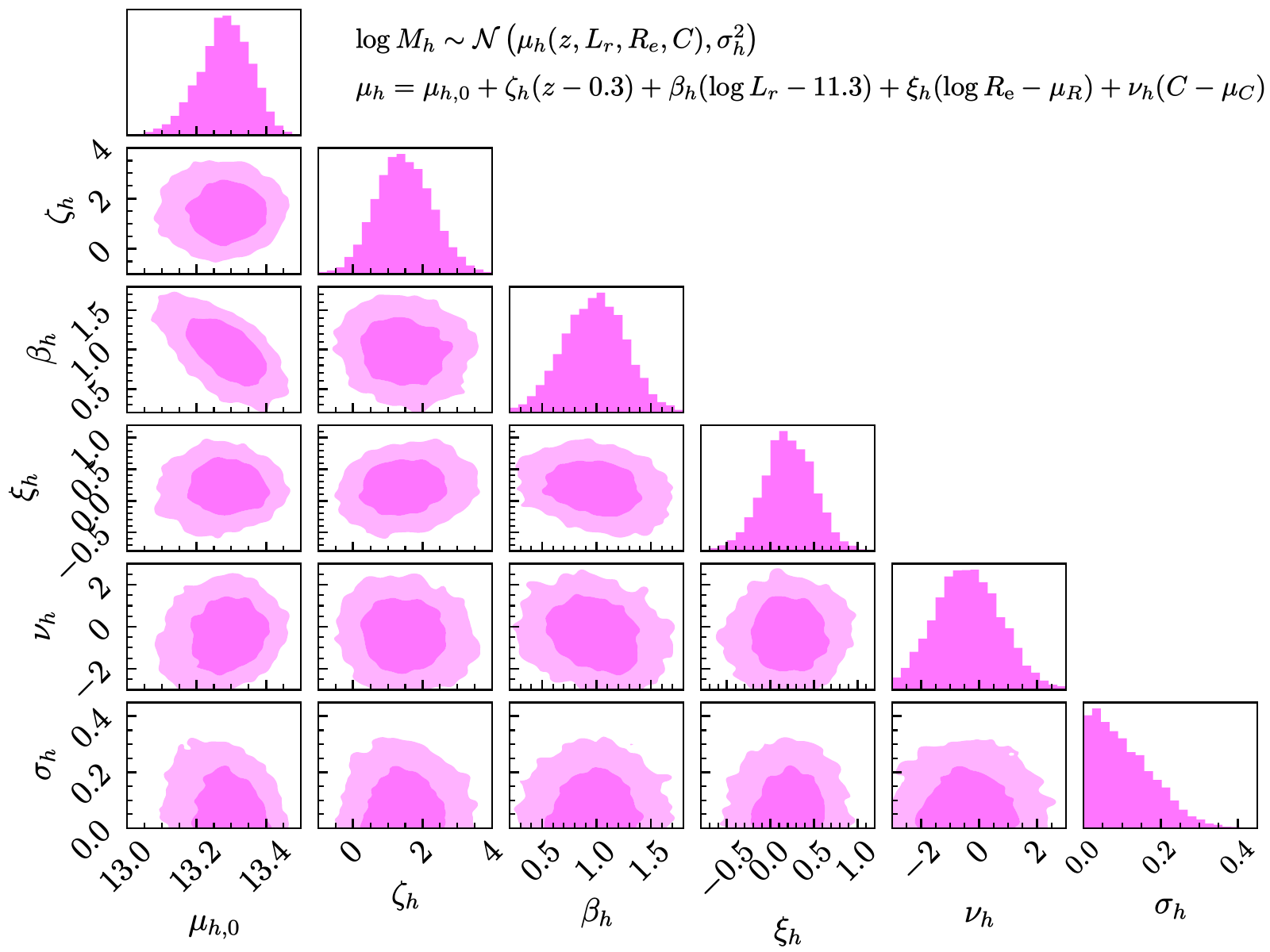}
\caption{
Posterior probability distribution of the model parameters describing the distribution in dark matter halo mass of the sample of lenses defined in \Eref{eq:dmdist}.
Contour levels correspond to a 68\% and 95\% enclosed probability.
\label{fig:dmcp}
}
\end{figure*}

The posterior probability distribution is shown in \Fref{fig:dmcp}. The median and 68\% credible region of the marginal posterior probability of each parameter is reported in \Tref{tab:dminfer}.
\begin{table*}
\caption{Parameters describing the halo mass distribution, \Eref{eq:hyperpars}.
}
\label{tab:dminfer}
\begin{tabular}{ccccl}
\hline
\hline
Parameter & Prior & Med.$\pm1\sigma$ & MAP & Description \\
\hline
$\muhzero$ & $U(11,15)$ & $13.28\pm0.07$ & 13.31 & Mean $\log{\mhalo}$ at $z=0.3$, $\log{\lumr}=11.3$, average size and colour. \\
$\zetah$ & $U(-3,3)$ & $1.4\pm0.8$ & 1.35 & Linear scaling of the average $\log{\mhalo}$ with redshift. \\
$\betah$ & $U(0,3)$ & $0.99\pm0.29$ & 1.02 & Linear scaling of the average $\log{\mhalo}$ with $\log{\lumr}$. \\
$\xih$ & $U(-1,1)$ & $0.20\pm0.29$ & 0.19 & Linear scaling of the average $\log{\mhalo}$ with excess size. \\
$\nuh$ & $U(-10,10)$ & $-0.4\pm1.2$ & -0.29 & Linear scaling of the average $\log{\mhalo}$ with excess colour. \\
$\sigmah$ & $U(0,3)$ & $0.09\pm0.08$ & 0.01 & Intrinsic scatter in $\log{\mhalo}$. 
\end{tabular}
\tablefoot{
Column (2): priors on the parameters.
Column (3): median and 68\% credible region of the marginal posterior probability distribution of each parameter given the data.
Column (4): maximum a posteriori value.
}
\end{table*}

We detect a positive correlation between halo mass and luminosity, with logarithmic slope $\betah=1.01\pm0.29$.
We do not detect correlations between halo mass and either redshift, size, or colour: parameters $\zetah$, $\xih$, and $\nuh$ are all consistent with $0$ within $2\sigma$.
We interpret these results as follows.

The correlation of halo mass with luminosity is closely related to that with stellar mass: this is confirmed by the fact that the value of $\betah$ that we measured is consistent with the slope of the halo mass-stellar mass correlation measured by \citet{Son++18} on a similarly selected sample of galaxies.
The lack of a strong correlation between halo mass and redshift at a fixed luminosity is the result of a combination of a few effects. First of all, there is an intrinsic evolution in the mapping between galaxy and halo masses, as both components grow with time; halo mass grows both via the accretion of matter and as a result of the evolution in the critical density of the Universe, which is an effect referred to as pseudo-evolution. Second, the luminosity evolution of passive galaxies implies that, at fixed luminosity, higher redshift galaxies are less massive than lower redshift ones, and therefore they live in less massive halos. Third, systematic errors in the photometric measurements of the background sources can introduce biases on the redshift evolution parameter $\zetah$. For example, if the source redshifts are systematically overestimated as a result of contamination from galaxies physically associated with the lens groups -- we only applied a global correction for this effect, which might not be entirely accurate -- this biases the inference of the halo mass of galaxies at a different redshift  differently due to the non-linear dependence of the lensing critical surface mass density on the lens and source redshift.
Our main motivation for introducing a scaling of halo mass with redshift is precisely to separate any evolution signal, whether real or artificial, from the dependence of halo mass on luminosity, size, and colour. Therefore, we do not investigate the nature of the $\zetah$ measurement  further.

The amplitude of the halo mass-size correlation, $\xih=0.17\pm0.27$, is consistent with the value found by \citet{SWB19} on a similarly selected sample of galaxies, $\xih=-0.14\pm0.17$, and corroborates the evidence against a strong correlation between these two quantities. Besides the data used, the main difference between the two measurements is that, in the present paper, $\xih$ describes a correlation at fixed redshift and luminosity; whereas in the case of \citet{SWB19}, $\xih$ refers to a correlation at a fixed stellar mass.
Our measurement of $\xih$ implies that the difference in the average halo mass between galaxies at the 84th and 16th percentile of the size distribution is 
\begin{equation}
\Delta_{2\sigmaR}\left<\log{\mhalo}\right> \equiv 2\sigmaR\xih = 0.07\pm0.10.
\end{equation}

Similarly,  given our constraint on the parameter describing the correlation between halo mass and colour, $\nuh$, we can quantify the difference in average halo mass between galaxies at the 84th and 16th percentile of the colour distribution at a fixed redshift, luminosity, and size, as
\begin{equation}
\Delta_{2\sigmaC}\left<\log{\mhalo}\right> \equiv 2\sigmaC\nuh = -0.03\pm0.11.
\end{equation}
The lack of a strong correlation between halo mass and rest-frame $u-g$ colour at a fixed redshift appears to contrast with the positive trend found by \citet{Z+M18} in their weak lensing and clustering study of red SDSS galaxies. 
We discuss the significance of this result in Section \ref{ssec:zu}.

Part of the signal captured by $\nuh$ could be due to a variation in the stellar mass-to-light ratio of galaxies with colour: if redder galaxies are intrinsically more massive at a fixed luminosity, their halo mass is also higher in virtue of the steep stellar mass-halo mass correlation. We discuss to what extent the coefficient $\nuh$ is sensitive to this effect in Section \ref{ssec:mltrend}.

Finally, we infer a value of the intrinsic scatter in $\log{\mhalo}$ of $\sigma_h = 0.08\pm0.08$.
While the inference is consistent with zero, our measurement does not allow us to put a strong constraint on this quantity: the 95\%-ile of the marginal posterior distribution in $\sigma_h$ is as large as $0.25$.

\subsection{Posterior predictive check}\label{ssec:pptest}

As a consistency test, we performed a posterior predictive check, which is how the goodness-of-fit is assessed with Bayesian hierarchical models.
Given our sample of lens galaxies, we drew realisations of the distribution of halo mass from the posterior probability distribution $\pr(\hyperp|\data)$. For each realisation, we predicted the weak lensing signal, which we then compared with the observed data.
In order to easily visualise the data, we considered the stacked weak lensing signal in bins for our comparison. 
We binned our lens sample in three different ways. First, we split it according to the galaxy excess size, dividing the sample into galaxies that lie above or below the average size given their redshift and luminosity, $\mu_R(z,\lumr)$. Second, we split the sample according to the colour between galaxies that are redder or bluer than the average colour given their properties, $\mu_C(z,\lumr,\reff)$. 
Third, we split the sample on the basis of the predicted average halo mass. In particular, we considered the set of model parameters $\hypermap$ that maximises the posterior probability distribution and used it to define the maximum a posteriori average halo mass, $\mu_h^{(\mathrm{MAP})}$. This is given by
\begin{equation}\label{eq:muhmap}
\begin{split}
\mu_h^{(\mathrm{MAP})} = & 13.31 + 1.35(z - 0.3) + 1.02(\log{\lumr} - 11.3) + \\
& 0.19\left(\log{\reff} - \mu_R(z,\lumr)\right) + \\ 
& -0.29\left(C - \mu_C(z,\lumr,\reff)\right).
\end{split}
\end{equation}
We then split the sample between galaxies with $\mu_h^{(\mathrm{MAP})}$ larger and smaller than the median value, which is $13.35$.

We computed the observed signal as follows. We made five radial bins in the range $50\,{\rm kpc} < R < 500\,{\rm kpc}$, then we converted the measurement of the tangential ellipticity of each source into an estimate of the excess surface mass density,
\begin{equation}\label{eq:deltasigma}
\Delta\Sigma^{\mathrm{(obs)}} \equiv \compellt\sigmacr
\end{equation}
and related uncertainty, where $\sigmacr$ is the critical surface mass density for gravitational lensing. Since we do not have exact source redshift measurements, we approximated $\sigmacr$ with an average over the source redshift distribution $n_z(z_s)$:
\begin{equation}
\sigmacr^{-1} \approx \frac{4\pi G}{c^2}D(\zlens)\int d\zsource n_w(\zsource) \frac{D(\zsource,\zlens)}{D(\zsource)},
\end{equation}
where $D(z)$ is the angular diameter distance as a function of redshift.

\begin{figure*}
\includegraphics[width=\textwidth]{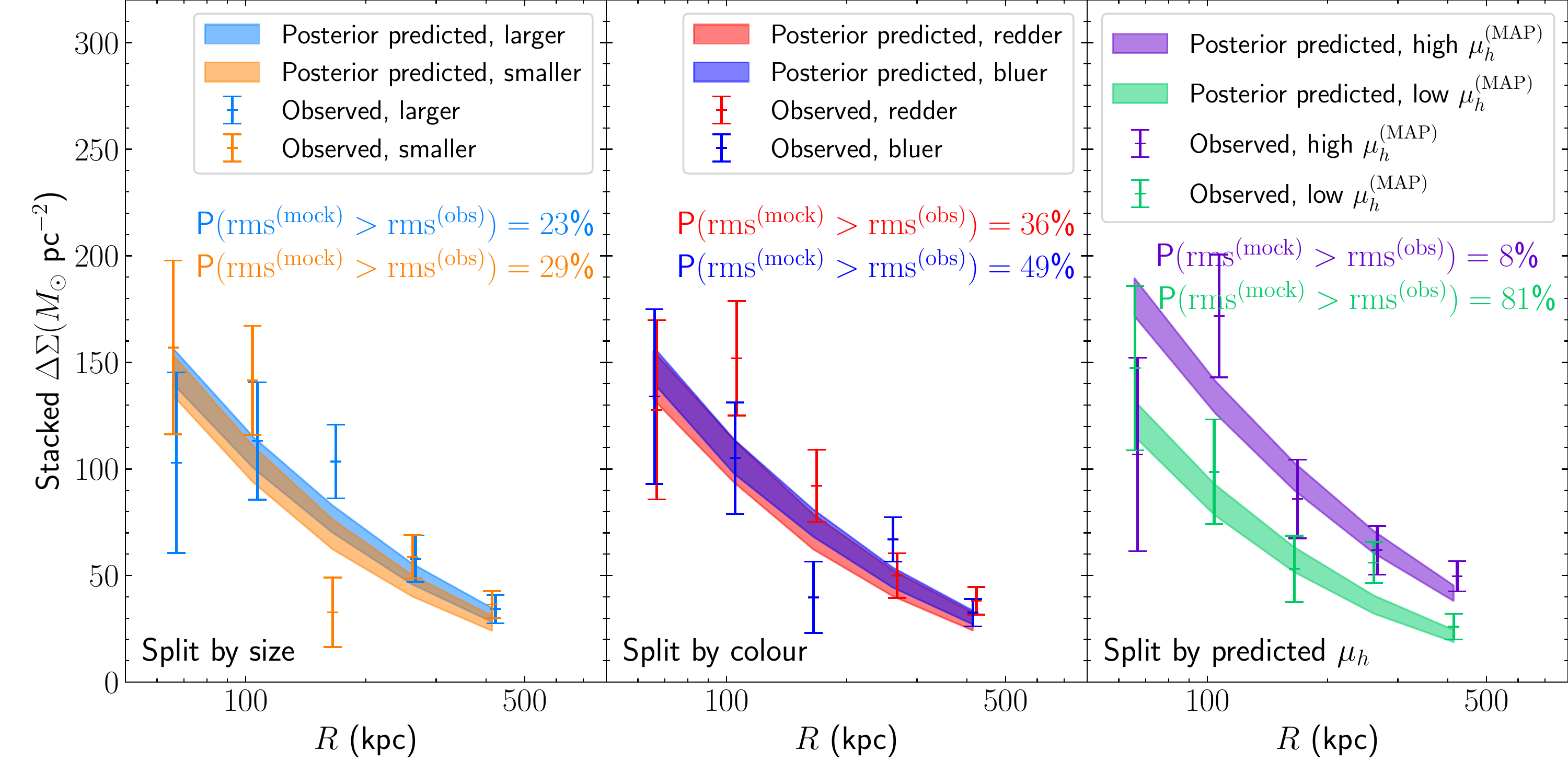}
\caption{
Stacked excess surface mass density profile of the lenses in our sample. The sample is split into two bins according to size (left panel), colour (middle panel), and predicted average halo mass, depending on the maximum a posteriori value of $\mu_h$ defined in \Eref{eq:muhmap} (right panel). 
The error bars show the excess surface mass density obtained by stacking the observed tangential ellipticity in five radial bins.
The bands show the same quantity obtained by posterior prediction from our model, without adding shape noise.
For each sub-sample, the figure reports the probability of obtaining a scatter around the posterior predicted $\Delta\Sigma$ profile larger than the observed value, under the assumption that the model describes the truth.
Small values for this quantity may indicate that the model is unable to reproduce these observations.
Error bars have been offset slightly in the horizontal direction, for clarity.
\label{fig:stacked}
}
\end{figure*}

The definition of \Eref{eq:deltasigma} is motivated by the fact that, in the case of a single lens, if $\Sigma(R)$ is the projected surface mass density of the lens and $\Sigma(R) \ll \sigmacr$, then the expectation value for the right-hand side quantity is the excess surface mass density $\Delta\Sigma(R)$ at the source position. The latter is defined as the difference between the average surface mass density within radius $R$ and $\Sigma(R)$:
\begin{equation}
\Delta\Sigma(R) = \bar{\Sigma}(<R) - \Sigma(R).
\end{equation}
For each radial bin, we obtained a weighted average of $\Delta\Sigma^{\mathrm{(obs)}}$, with an inverse variance weighting scheme, and related uncertainty. 
The result is shown as error bars in \Fref{fig:stacked}.

We subsequently obtained the posterior-predicted expectation value of the stacked excess surface density profile, $\Delta\Sigma^{(\mathrm{pp})}$, as follows. At each draw of the values of the model parameters from the posterior probability distribution, we randomly drew a value for the halo mass and concentration for each lens in the sample. Then we computed the expectation value of the tangential ellipticity of each source, given its position, redshift, and the mass density profile of its lens. Finally, we obtained the inverse-variance weighted $\Delta\Sigma$, as was done with the observed data. The 16th and 84th percentile of the posterior-predicted $\Delta\Sigma^{(\mathrm{pp})}$ profile of the two lens sub-samples for each binning scheme are shown as bands in \Fref{fig:stacked}.

The $\Delta\Sigma^{(\mathrm{pp})}$ profile reproduces the average observed signal for each sub-sample. 
In order to quantify the goodness of fit, we considered the standard deviation in the observed $\Delta\Sigma$ around the posterior predicted profile as test statistics: assuming our model to be accurate, we estimated the probability of obtaining a higher standard deviation than the observed value. Very low values for this probability, 5\% or lower, might indicate the inability of the model to reproduce these data.

For each posterior draw, we generated mock $\Delta\Sigma$ observations in the five radial bins by adding noise to the posterior predicted profile; computed the standard deviation of the mock data points around $\Delta\Sigma^{(\mathrm{pp})}$, $\mathrm{rms}^{(\mathrm{mock})}$; and compared this quantity with the standard deviation of the observed data points around $\Delta\Sigma^{(\mathrm{pp})}$, $\mathrm{rms}^{\mathrm{(obs)}}$. 
In \Fref{fig:stacked} we report the probability that $\mathrm{rms}^{(\mathrm{mock})} > \mathrm{rms}^{(\mathrm{obs})}$ for each sub-sample.
The smallest value of this probability is $8\%$, which occurs for the sub-sample of higher $\mu_h^{\mathrm{(MAP)}}$. This means that there is no strong reason to question the ability of our model to reproduce the stacked weak lensing data.

\section{Distribution as a function of robust observables}\label{sect:robust}

In this section we fit for the distribution of halo mass of the sample as a function of redshift; rest-frame $r$-band luminosity within an aperture of $10$~kpc, $\lten$; light-weighted projected slope within $10$~kpc (defined in \Eref{eq:slopedef}), $\slope$; and the rest-frame $u-g$ colour, $C$.
Compared to the analysis of \Sref{sect:halo}, we replace the total luminosity and the half-light radius with $\lten$ and $\slope$. The motivation for this is to obtain a description of the halo mass distribution that is robust with respect to assumptions on the asymptotic behaviour of the surface brightness profile of the lens galaxies.

Similarly to our approach in \Sref{sect:indep}, we first fit for the distribution in colour as a function of redshift, $\lten$ and $\slope$. We do this so that we can later model the distribution in halo mass as a function of excess colour, that is the difference between the colour of a galaxy and the average colour of galaxies with the same redshift, $\lten$ and $\slope$.
We model the colour distribution with the same functional form as \Eref{eq:colourdist}, but now expressed as a function of the new set of parameters:
\begin{equation}\label{eq:newcolourdist}
\mathcal{K}(C|z,\lten,\slope) = \frac{1}{\sqrt{2\pi}\sigmaC}\exp{\left\{-\frac{\left(C - \mu_C(z,\lten,\slope)\right)^2}{2\sigmaC^2}\right\}}.
\end{equation}
The mean colour $\muC$ is allowed to scale with $z$, $\lten$, and $\slope$ as follows:
\begin{equation}\label{eq:newmeancolour}
\begin{split}
\muC(z,\lten,\slope) = & \muCzero + \zetaC(z - 0.3) + \lambdaC(\log{\lumr} - 10.9) + \\ & \gammaC\left(\slope - 1.5\right).
\end{split}
\end{equation}

We fitted for the parameters of the distribution of \Eref{eq:newcolourdist}: the results are reported in \Tref{tab:newcolourfit}.
Galaxy colour does not depend on $\lten$ or $\slope$ at a fixed redshift.
This result matches the finding of \Sref{sect:indep}: the colour of these galaxies is independent of their surface brightness profile.
At a fixed surface brightness profile, colour decreases with increasing redshift with the same trend previously measured in \Sref{sect:indep}: $\zetaC=-0.418\pm0.016$.

\begin{table*}
\caption{Parameters describing the distribution in colour as a function of redshift, $\lten$ and $\slope$.
}
\label{tab:newcolourfit}
\begin{tabular}{cccl}
\hline
\hline
Parameter & Prior & Med.$\pm1\sigma$ & Description \\
\hline
$\muCzero$ & $U(1,3)$ & $1.751\pm0.001$ & Mean $u-g$ colour at $z=0.3$, $\log{\lten}=10.9$ and average size. \\
$\zetaC$ & $U(-10,10)$ & $-0.418\pm0.017$ & Linear scaling of the average colour with redshift. \\
$\betaC$ & $U(-3,3)$ & $0.005\pm0.010$ & Linear scaling of the average colour with $\log{\lten}$. \\
$\xiC$ & $U(-10,10)$ & $0.009\pm0.010$ & Linear scaling of the average colour with $\slope$. \\
$\sigmaC$ & $U(0,1)$ & $0.053\pm0.001$ & Intrinsic scatter in colour at fixed redshift, $\lten$ and $\slope$. \\\end{tabular}
\tablefoot{
The parameters are introduced in \Eref{eq:newcolourdist} and \Eref{eq:newmeancolour}.
Column (2): priors on the parameters.
Column (3): median and 68\% credible region of the marginal posterior probability distribution of each parameter given the data.
}
\end{table*}

We then model the distribution in halo mass and concentration of the sample.
We use the same functional form as \Eref{eq:dmdist}, expressed in terms of the new set of variables, for this purpose:
\begin{equation}\label{eq:halodistrobust}
\begin{split}
\pr(\log{\mhalo},\log{\chalo}|z,\lten,\slope,C) = & \mathcal{H}(\log{\mhalo}|z,\lten,\slope,C)\times \\
& \mathcal{C}(\log{\chalo}|\mhalo).
\end{split}
\end{equation}
Similarly to \Eref{eq:mhalodist}, the term $\mathcal{H}$ is a Gaussian in $\log{\mhalo}$, with a mean $\mu_h$ that is given by
\begin{equation}\label{eq:muhalorobust}
\begin{split}
\mu_h(z,\lten,\slope,C) = & \mu_{h,0} + \zetah(z - 0.3) + \lambdah(\log{\lumr} - 10.9) + \\
& \gamma_h\left(\slope - 1.5\right) + \\ 
& \nuh\left(C - \mu_C(z,\lten,\slope)\right),
\end{split}
\end{equation}
and dispersion $\sigma_h$.
The quantity $\muC(z,\lten,\slope)$ is the average colour given by \Eref{eq:newmeancolour}, with the coefficients $\zetaC$, $\lambda_C$, and $\gammaC$ fixed to the median marginal posterior values reported in \Tref{tab:newcolourfit}.
The concentration distribution $\mathcal{C}$ is the same as \Eref{eq:c200dist}.

\begin{figure*}
\includegraphics[width=\textwidth]{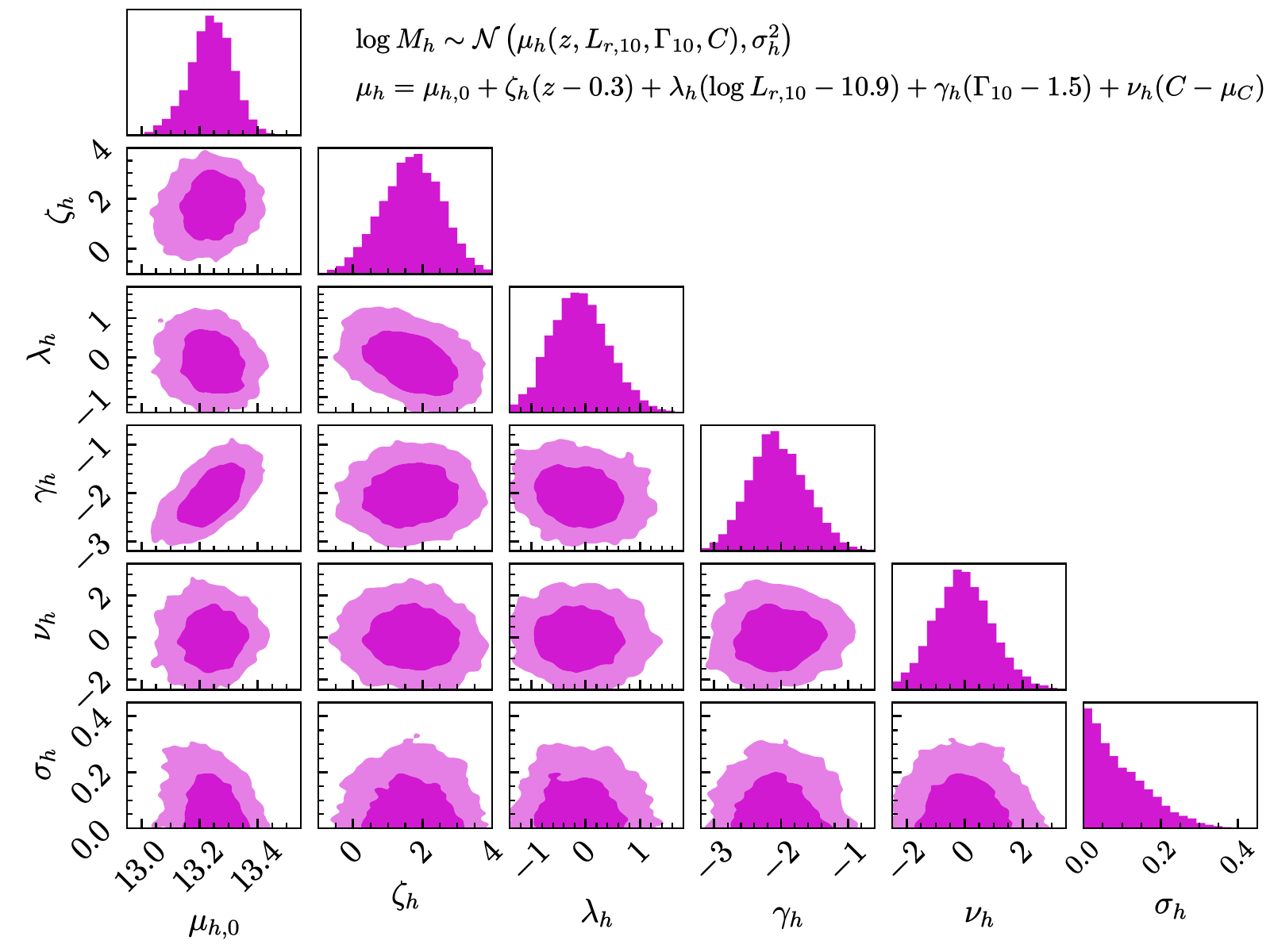}
\caption{
Posterior probability distribution of the model parameters describing the distribution in halo mass as a function of redshift, $\lten$ and $\slope$, \Eref{eq:halodistrobust}.
Contour levels correspond to a 68\% and 95\% enclosed probability.
\label{fig:robustcp}
}
\end{figure*}

In \Fref{fig:robustcp} we show the posterior probability distribution of the model parameters $(\muhzero,\zetah,\lambdah,\gammah,\nuh,\sigmah)$. The median and 68\% credible region of the marginal posterior of each parameter is given in \Tref{tab:robusthalo}.
Our measurement of $\lambdah$ does not let us determine whether halo mass increases or decreases with $\lten$ at a fixed redshift. However, we find that halo mass increases with decreasing $\slope$. The values of the parameters describing the redshift and colour dependence of the halo mass at a fixed surface brightness profile, $\zetaC$ and $\nuC$, are slightly different compared to the inference obtained in \Sref{sect:halo}, but consistent within the uncertainty.
The difference in the average halo mass between galaxies at the 84\%-ile and 16\%-ile of the colour distribution that we obtain in this case is
\begin{equation}
\Delta_{2\sigmaC}\left<\log{\mhalo}\right> \equiv 2\sigmaC\nuh = 0.00\pm0.11.
\end{equation}
This indicates that our constraint on the correlation between halo mass and colour is somewhat model-dependent, but, given the current uncertainty level, it is still robust with respect to the two parameterisations of the surface brightness profile that we employed.
Finally we point out that, although the value of the parameter $\muhzero$ measured here differs from the value found in \Sref{sect:halo}, this is simply the consequence of adopting a different pivot point for the description of the mean halo mass of \Eref{eq:meanhalo}, compared to \Eref{eq:muhalorobust}.

\begin{table*}
\caption{Parameters describing the distribution in halo mass as a function of redshift, $\lten$ and $\slope$, \Eref{eq:halodistrobust}.
\label{tab:robusthalo}
}
\begin{tabular}{ccccl}
\hline
\hline
Parameter & Prior & Med.$\pm1\sigma$ & MAP & Description \\
\hline
$\muhzero$ & $U(11,15)$ & $13.24\pm0.07$ & 13.27 & Mean $\log{\mhalo}$ at $z=0.3$, $\log{\lten}=10.9$, $\slope=1.5$ and average colour. \\
$\zetah$ & $U(-10,10)$ & $1.7\pm0.9$ & 1.70 & Linear scaling of the average $\log{\mhalo}$ with redshift. \\
$\lambdah$ & $U(-10,10)$ & $-0.1\pm0.5$ & -0.08 & Linear scaling of the average $\log{\mhalo}$ with $\log{\lten}$. \\
$\gammah$ & $U(-10,10)$ & $-2.1\pm0.4$ & -2.06 & Linear scaling of the average $\log{\mhalo}$ with $\slope$. \\
$\nuh$ & $U(-10,10)$ & $-0.1\pm1.0$ & 0.08 & Linear scaling of the average $\log{\mhalo}$ with excess colour. \\
$\sigmah$ & $U(0,3)$ & $0.08\pm0.08$ & 0.00 & Intrinsic scatter in $\log{\mhalo}$. \\\end{tabular}
\tablefoot{
Column (2): priors on the parameters.
Column (3): median and 68\% credible region of the marginal posterior probability distribution of each parameter given the data.
Column (4): maximum a posteriori value.
}
\end{table*}

\section{Discussion}\label{sect:discuss}

\subsection{Expectations from variations in $M_*/L$}\label{ssec:mltrend}

The analysis of \Sref{sect:halo} and \Sref{sect:robust} shows that, at a fixed redshift and surface brightness distribution, there is not a large difference in the average halo mass of redder and bluer elliptical galaxies.
If the stellar mass-to-light ratio was identical for all of our galaxies, expressing trends in terms of luminosity or stellar mass would be equivalent.
However, colour differences reflect differences in the properties of the stellar population, and therefore we can expect the stellar mass-to-light ratio to vary across our sample.
In this section we investigate to what extent our result can change when using stellar mass in place of luminosity. 

We proceed as follows.
We use composite stellar population synthesis models obtained with {\sc Galaxev} \citep{BC03} to assign stellar masses to galaxies.
Rather than fitting for the stellar population parameters to the full set of photometric data, however, we assume a simplified model that results in a one-to-one mapping between the rest-frame $u-g$ colour and stellar mass-to-light ratio.
We do this to maximise the correlation between these two properties, so as to obtain an upper limit on the effect of variations in mass-to-light ratio with colour.
Then, we repeat the analysis carried out in \Sref{sect:halo}, using these assigned stellar masses in place of the luminosities.

\begin{figure}
\includegraphics[width=\columnwidth]{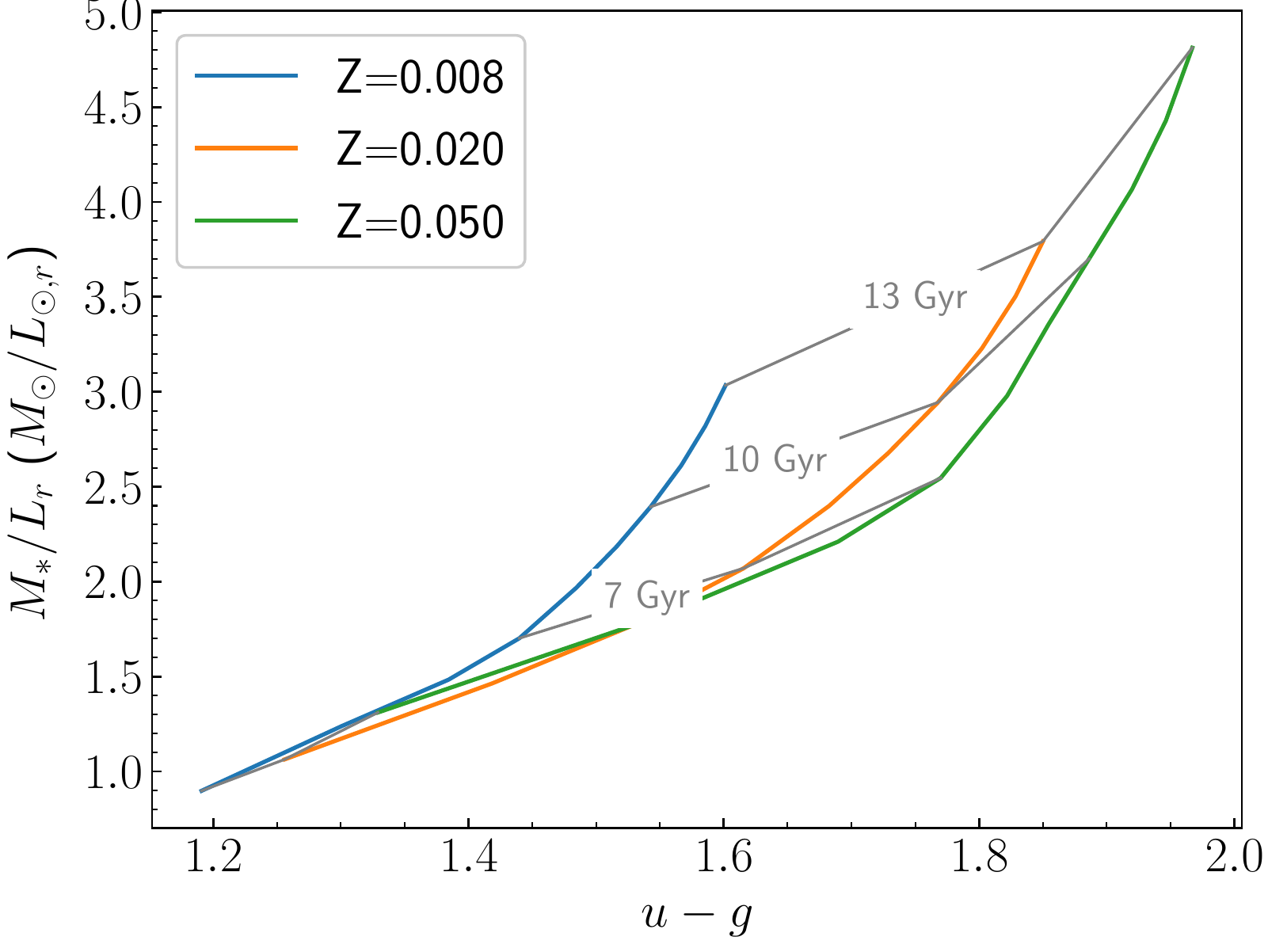}
\caption{
Stellar mass-to-light ratio as a function of rest-frame $u-g$ colour, for three composite stellar population models obtained with {\sc Galaxev} \citep{BC03}.
Each model has a fixed metallicity and an exponentially declining star formation history with a decay time of $1$~Gyr. Each curve was obtained by varying the time since the initial burst. The grey lines connect points of fixed time since the initial burst, as indicated by the labels.
The model corresponding to a metallicity $Z=0.05$ is used in Section \ref{ssec:mltrend} to assign stellar masses to galaxies, given their observed colour.
\label{fig:mlmodel}
}
\end{figure}

We assume a model with a fixed metallicity of $Z=0.05$ and an exponentially decaying star formation history with decay time $\tau=1$~Gyr. The time since the initial burst is the only parameter that we leave free to vary.
\Fref{fig:mlmodel} shows how the stellar mass-to-light ratio varies as a function of colour for this model (green curve). Given the value of $u-g$ of each galaxy, we use this curve to assign a value of $M_*/L_r$ and, consequently, a stellar mass to it.
For reference, in \Fref{fig:mlmodel} we also show curves for two different values of the metallicity covered by the \citet{BC03} stellar population library.
Although the value of the metallicity that we adopted is relatively high, we made this choice to ensure that our model can span the whole range of the observed $u-g$ colour.

Following \Sref{sect:indep}, we first determined the distribution in half-light radius as a function of the redshift and stellar mass of the sample. We found the following expression for the average $\log{\reff}$:
\begin{equation}
\muR(z,\mstar) = 1.08 -0.21(z - 0.3) + 1.31(\log{\mstar} - 11.6),
\end{equation}
and an intrinsic scatter in $\log{\reff}$ at a fixed $z$ and $\mstar$ of $\sigmaR = 0.20$.
Then we measured the distribution in colour as a function of redshift, stellar mass, and excess size.
This has a best-fit mean given by
\begin{equation}
\begin{split}
\muC(z,\mstar,\reff) = & 1.74 - 0.40(z-0.3) + 0.05(\log{\mstar} - 11.6) + \\
& - 0.10(\log{\reff} - \muR(z,\mstar)),
\end{split}
\end{equation}
and a scatter of $\sigmaC=0.05$.
We point out that, as we anticipated in \Sref{sect:indep}, colour has a non-negligible anti-correlation with excess size, at fixed stellar mass and redshift.

We fitted for the distribution in halo mass as a function of redshift, stellar mass, excess size, and excess colour. We used the same method and the same model as in \Sref{sect:halo}, with stellar mass in place of the luminosity. In particular, we assumed a Gaussian distribution in $\log{\mhalo}$, with the following mean,
\begin{equation}\label{eq:muhalomstar}
\begin{split}
\muh(z,\mstar,\reff,C) = & \muhzero + \zetah(z - 0.3) + \betah(\log{\mstar} - 11.6) + \\
& \xih\left(\log{\reff} - \muR(z,\mstar)\right) + \\ 
& \nuh\left(C - \muC(z,\mstar,\reff)\right),
\end{split}
\end{equation}
and dispersion $\sigmah$.

\begin{figure}
\includegraphics[width=\columnwidth]{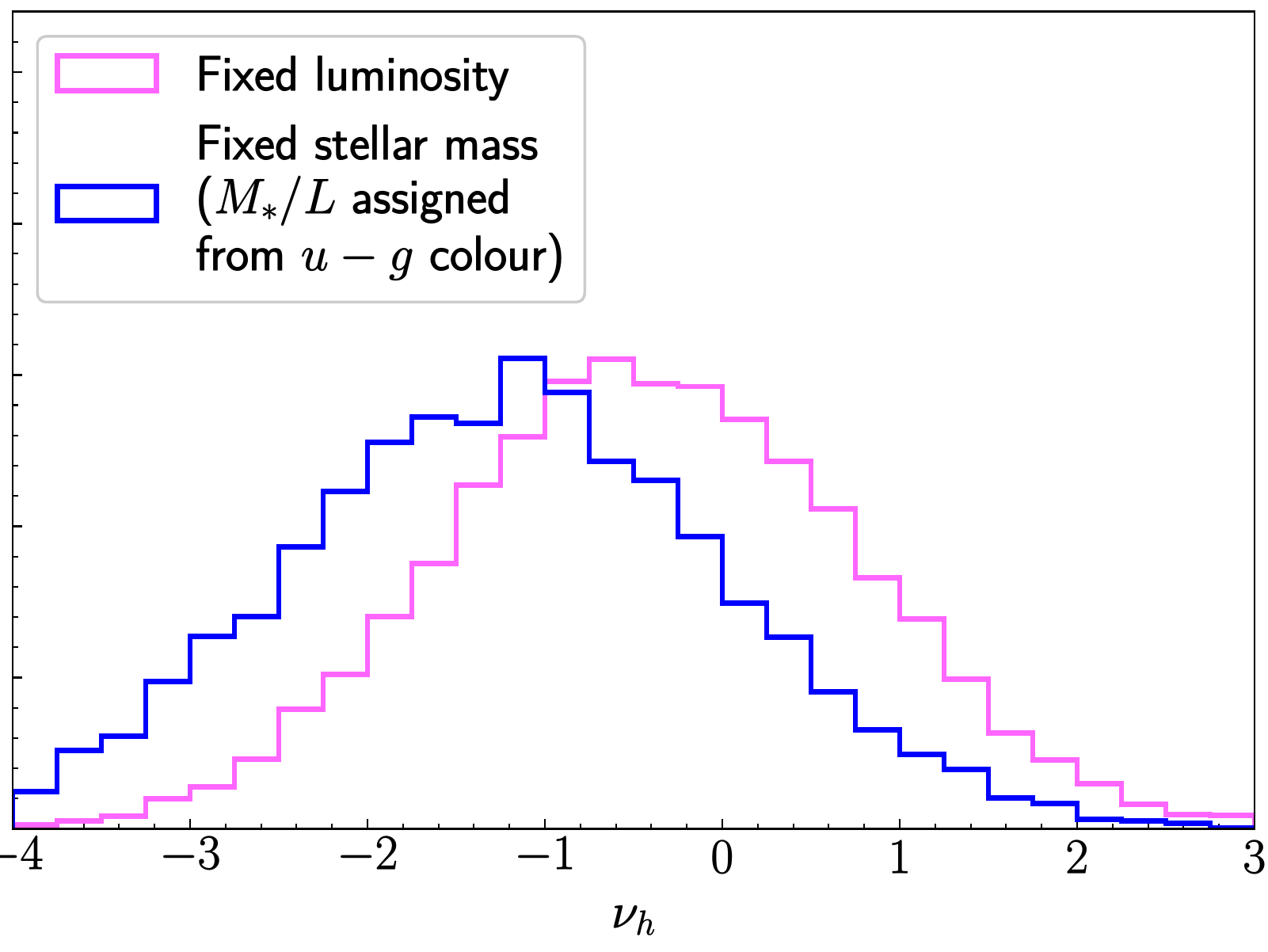}
\caption{
Colour dependence of the halo mass.
Blue histogram: Marginal posterior probability distribution of the parameter $\nuh$, describing the dependence  of the average halo mass as a function of excess colour at a fixed redshift, stellar mass, and size, as defined in \Eref{eq:muhalomstar}.
Pink histogram: Marginal posterior probability distribution of the parameter $\nuh$, describing the dependence of the average halo mass as a function of excess colour at a fixed luminosity and size, as inferred in \Sref{sect:halo}.
\label{fig:mstarnuh}
}
\end{figure}

In \Tref{tab:mstar} we report the median and 68\% credible region of the marginal posterior probability distribution of each model parameter. \Fref{fig:mstarnuh} shows the marginal posterior of the parameter $\nuh$, which in the context of this analysis quantifies the correlation between halo mass and colour at a fixed stellar mass, redshift, and size (blue histogram). The same figure shows the value of $\nuh$ obtained from the analysis of \Sref{sect:halo} (pink histogram).
\begin{table*}
\caption{Parameters describing the distribution in halo mass as a function of redshift, stellar mass, half-light radius, and colour, \Eref{eq:muhalomstar}.
\label{tab:mstar}
}
\begin{tabular}{cccl}
\hline
\hline
Parameter & Prior & Med.$\pm1\sigma$ & Description \\
\hline
$\mu_{h,0}$ & $U(11,15)$ & $13.18\pm0.09$ & Mean $\log{\mhalo}$ at $z=0.3$, $\log{\mstar}=11.6$, average size and colour. \\
$\zeta_h$ & $U(-3,3)$ & $1.6\pm0.8$ & Linear scaling of the average $\log{\mhalo}$ with redshift. \\
$\beta_h$ & $U(0,3)$ & $0.99\pm0.26$ & Linear scaling of the average $\log{\mhalo}$ with $\log{\mstar}$. \\
$\xi_h$ & $U(-1,1)$ & $0.17\pm0.28$ & Linear scaling of the average $\log{\mhalo}$ with excess size. \\
$\nu_h$ & $U(-10,10)$ & $-1.2\pm1.2$ & Linear scaling of the average $\log{\mhalo}$ with excess colour. \\
$\sigma_h$ & $U(0,3)$ & $0.09\pm0.08$ & Intrinsic scatter in $\log{\mhalo}$. 
\end{tabular}
\tablefoot{
Column (2): priors on the parameters.
Column (3): median and 68\% credible region of the marginal posterior probability distribution of each parameter given the data.
}
\end{table*}

Compared to the analysis of \Sref{sect:halo}, the inferred value of $\nuh$ is lower, but still consistent with no correlation: $\nuh=-1.2\pm1.2$.
This result was obtained under the assumption that the stellar mass-to-light ratio has a perfect positive correlation with colour, and therefore represents an extreme case.
The true amplitude of the halo mass-colour correlation at a fixed stellar mass is somewhere between the value obtained using luminosity and that reported in this section.
In order to further clarify this issue, however, it is necessary to obtain robust estimates of galaxy stellar mass-to-light ratios. As we explained in Section \ref{sect:intro}, this is currently very challenging.

\subsection{Possible sources of systematic errors}\label{ssec:systematics}

Our work consists of measuring the distribution of halo mass as a function of various galaxy properties.
We put particular care in ensuring that these properties, including redshift, surface brightness profile, and colour, were measured robustly.
Nevertheless, some of these measurements could still be affected by systematic errors.

One possible source of error is the presence of colour gradients in the galaxies in our sample. Massive galaxies are known to have, on average, negative colour gradients \citep[that is, they are redder in the inner parts, see for example][]{Tor++10,Szo++13}; however, we implicitly assumed spatially uniform colours when deriving sizes, luminosities, and rest-frame colours.
This implies that our measurements are sensitive to the choice of the photometric band used for the S\'{e}rsic profile fits, the $r-$band, and the method used to measure colours, which relies on the GAaP method of \citet{Kui++15}.

A negative colour gradient means that the size of a galaxy is smaller when measured at longer wavelengths. This can introduce a spurious correlation between size and redshift because the observed-frame $r-$band corresponds to longer wavelengths for galaxies at higher $z$. Such a correlation is absorbed by the parameter $\zeta_\mathrm{R}$ of our model, which describes the scaling of size with redshift (see \Eref{eq:meansize}).
However, if gradients are a strong function of luminosity, then that would alter the distribution of galaxies in luminosity-size space, potentially introducing a bias in our estimate of parameters $\betah$ and $\xih$.
Similarly, our measurement of the trend of halo mass with colour is robust with respect to gradients, as long as these gradients scale with redshift, luminosity, and size in a way that is captured by the model distribution of \Eref{eq:meancolour}.
However, if gradients correlate with halo mass in a way that is not captured by our model, then biases are introduced.

The correct approach to dealing with colour gradients would be to adopt a more flexible model for the surface brightness distribution of galaxies, with a definition for galaxy size that is independent of the photometric band used, and with the presence of gradients. The model for the halo mass distribution, then, would need to allow for a scaling of halo mass with the colour gradient of a galaxy. That, however, is beyond the scope of this work.

Another possible source of systematic error is the assumption of an NFW density profile with a mass-concentration relation for the dark matter on which our measurement of the halo mass distribution is based. 
The choice of the density profile is degenerate with the halo mass: for example, assuming a lower concentration leads to a higher inference on $\mhalo$ \citep[see Section 5.3 of][]{Tay++20}.
In principle, then, a positive trend between halo mass and a galaxy observable could escape our detection if this observable also has an anti-correlation with concentration.
It is then important, when comparing our measurements with model predictions, to define the halo masses of the model in as similar a way as possible to ours: for example, by fitting the projected surface mass density profile in the radial range $50-500$~kpc with an NFW profile and a mass-concentration relation given by \Eref{eq:c200mu}.

We also assumed a spherical distribution for the dark matter mass. This is, on average, a good assumption, but individual halos are close to elliptical in projection \citep[see for example][]{Geo++21,Sch++21}.
To first approximation, ignoring ellipticity worsens the goodness of fit of the lensing data around individual lenses. Our model can compensate for this effect by increasing the intrinsic scatter in the halo mass.

We interpreted the lensing signal under the assumption that all of the galaxies in the sample are at the centre of their host halo.
Although we removed probable satellite galaxies with the use of the redMaPPer cluster catalogue, there could still be residual satellites in the sample due to the incompleteness of the catalogue.
Nevertheless, \citet{S+L18} show that the inference method on which our work is based is robust with respect to the presence of satellites: for example, a satellite fraction as large as 16\% causes a negligible bias on the inferred average halo mass.

The stellar contribution to the lensing signal was estimated by using stellar masses obtained via stellar population synthesis modelling. As we discussed above, these stellar masses are subject to systematic uncertainties.
We repeated the analysis by increasing the stellar masses by $0.25$~dex, which corresponds to the difference in stellar mass-to-light ratio between a Salpeter and a Chabrier IMF. The main change that we found is that the average halo mass parameter $\muhzero$ decreases by $0.05$, compared to the fiducial analysis.

Finally, we ignored the contribution of gas to the total mass.
This means that, implicitly, the gas is described by the dark matter component, in our model. We expect the fraction of gas to total mass to be smaller than $10\%$ and roughly constant with halo mass \citep{D+R15}, and therefore it is not significant for the results of our study.

\subsection{Comparison with the \citet{Z+M18} study}\label{ssec:zu}

\citet{Z+M18}, using weak lensing and galaxy clustering, report a positive trend between halo mass and rest-frame $g-r$ colour at a fixed stellar mass for a sample of colour-selected red galaxies from SDSS.
They also show how their measurement favours a halo-quenching model, a scenario in which the probability of finding a quenched (that is, red) galaxy in a given halo primarily correlates with the halo mass, as opposed to a model in which the galaxy colour primarily depends on the halo assembly time.
Confirming or questioning the trend with our independent measurement can therefore have important implications for galaxy evolution models.

\citet{Z+M18} do not explicitly report a distribution of halo mass as a function of colour, and therefore it is difficult to make a direct comparison with our results.
Nevertheless, their Figure 7 shows the weak lensing signal, $\Delta\Sigma$, measured in three stellar mass bins for two subsets of galaxies split on the basis of their $g-r$ colour.
Their data show that, in their highest stellar mass bin, the values of $\Delta\Sigma$ for redder galaxies are $\sim50\%$ higher than those of bluer galaxies over the radial range covered by our measurements.
At a fixed physical distance from the lens centre and fixed halo concentration, $\Delta\Sigma$ roughly scales with a power $0.6-0.7$ of the halo mass. Therefore, a 50\% difference in $\Delta\Sigma$ corresponds to a halo mass difference of $0.3$~dex.
Our posterior prediction on the difference in average halo mass between our redder and bluer sub-samples rules out such a large value with more than 99\% probability, and the discrepancy becomes even larger when using stellar mass in place of luminosity to split the sample. Therefore, our measurements are in strong contrast with those of \citet{Z+M18}.

There are, however, several differences between our analysis and that of \citet{Z+M18}. Firstly, \citet{Z+M18} used $g-r$ to distinguish among galaxies of a different colour, while we used $u-g$. However, for red galaxies, such as those in our sample, the two colours are very strongly correlated, and therefore our results are robust with respect to the particular choice of photometric bands. 
More important differences are the selection criteria and the stellar mass measurements used: 
we drew galaxies from the Luminous Red Galaxy sample of SDSS, while \citet{Z+M18} used the Main Spectroscopic Sample \citep{Str++02}.
Although there is overlap between the two samples, the latter is dominated by lower-mass and lower-redshift galaxies.
Moreover, we applied a morphological selection and obtained luminosities from S\'{e}rsic profile fitting to KiDS data, while \citet{Z+M18} used colour alone to define their sample of red galaxies and relied on stellar masses obtained from SDSS. 

In light of these differences between the two studies, there can be several possible explanations for the origin of the discrepancy. 
First, the stellar masses used in one of the two studies could be biased in a colour-dependent way: since stellar mass scales with halo mass, any stellar mass-colour correlation would affect the estimate of $\nuh$ (as Section \ref{ssec:mltrend} shows).

Second, a selection based purely on colour, such as that of \citet{Z+M18}, results in a sample of red galaxies with a non-negligible fraction of lenticular galaxies, bulge-dominated galaxies with spiral arms, or galaxies with an irregular morphology. If halo mass is a function of morphology, such a dependence could show up as a correlation with colour. Such a correlation could also be observed if the stellar mass measurements are biased in a morphology-dependent way. This could happen, for example, if the presence of spiral arms biases the total flux of a galaxy obtained by fitting a simply parameterised surface brightness profile \citep{Son21}.
We tested this hypothesis by repeating our analysis while relaxing our morphology selection criterion. This resulted in the inclusion of an additional 157 galaxies, representing about $6\%$ of the sample. 
The median value of the marginal posterior on $\nuh$ changed from $\nuh=-0.3$ to $\nuh=0.0$: the effect of including disk galaxies increases the correlation between halo mass and colour, but not to the level required to explain the discrepancy with the \citet{Z+M18} measurement.
The fraction of red galaxies with a disky morphology, however, is a strong function of stellar mass. Therefore, the morphological selection could still be important in a sample such as that of \citet{Z+M18}.

Third, the inference method employed here  substantially differs from that of \citet{Z+M18}: while we fit our model directly to the individual shape measurements around each lens, \citet{Z+M18} first binned lenses and then fitted the stacked weak lensing data.
In order to understand the impact that binning can have on the measurement, we repeated our analysis by splitting the lens sample into two bins according to colour, using the same procedure as in section \ref{ssec:pptest}, and then fitting for the average halo mass of each sub-sample separately. 
Since information on the diversity of galaxies within a bin is usually lost in a binned analysis, we did not account for a dependence of the halo mass on any galaxy property: in other words, we set the model parameters $\zeta_{\mathrm{h}}$, $\beta_{\mathrm{h}}$, $\xi_{\mathrm{h}}$, and $\nu_{\mathrm{h}}$ to zero.

We found a value of $\mu_{\mathrm{h},0} = 13.38\pm0.08$ for the redder sub-sample and $\mu_{\mathrm{h},0} = 13.26\pm0.10$ for the bluer sub-sample.
This result tells us that, if we ignore correlations between halo mass and galaxy properties within a bin, our data favour a model in which redder galaxies live in more massive halos on average; although, solutions in which the two sub-samples share the same halo mass are well within $2\sigma$.
The fact that the full model instead favours a slightly negative value of $\nu_h$ indicates that the apparent difference in average halo mass found in the binned analysis is more likely the result of correlations between halo mass and galaxy properties other than colour, and most importantly luminosity.
Nevertheless, this experiment suggests that a difference in the inferred average halo mass of the order of $0.1$~dex with respect to the \citet{Z+M18} study could be due to the choice of binning the sample.

Fourth, the colour dependence of halo mass could be a strong function of stellar mass disappearing at the very high masses probed by our sample. 
Lastly, there could also be biases in the colour or weak lensing measurements in either of the two studies.

In order to rule out any of these hypotheses, it would be necessary to repeat our analysis with the same selection criteria, stellar mass measurements, and colour measurements as those used by \citet{Z+M18}. That, however, is beyond the scope of this work.

\section{Conclusions}\label{sect:concl}

We carried out a study of the dark matter halo mass distribution of a sample of massive elliptical galaxies, selected from a combination of spectroscopic and morphological criteria.
We first measured the distribution in $\log{\mhalo}$ as a function of redshift, luminosity, half-light radius, and colour. 
Then we repeated the analysis by describing the surface brightness distribution of galaxies in terms of the luminosity within a physical aperture of $10$~kpc, $\lten$, and the light-weighted projected slope of the surface brightness profile within the same aperture, $\slope$.
The first description allows for a more intuitive interpretation of the results. The latter is more robust with respect to assumptions on the surface brightness profile of the galaxies.

Our measurements can be used to test theoretical models for the distribution of dark matter halo masses of massive elliptical galaxies, provided that those models can make predictions on galaxy stellar mass-to-light ratios.
For models that can predict the full radial profile of the stellar distribution of galaxies, such as hydrodynamical simulations, we recommend using our description of the halo mass distribution in terms of $\lten$ and $\slope$, presented in \Sref{sect:robust}, as a reference. This would allow one to avoid introducing biases related to the extrapolation of the S\'{e}rsic surface brightness model out to very large radii, where it is not constrained by the data.

Our analysis ruled out strong correlations between halo mass and galaxy size at a fixed luminosity, corroborating previous findings from \citet{SWB19}.
In particular, the difference in the average halo mass between galaxies at the 84\%-ile and 16\%-ile of the distribution in half-light radius at a fixed luminosity and redshift is $0.06\pm0.11$~dex (68\% credible region).
We also ruled out a strong correlation between halo mass and rest-frame $u-g$ colour.
The value of the coefficient describing the halo mass-colour correlation slightly depends on whether we adopted a description in terms of luminosity and size, or $\lten$ and $\slope$. In the latter case,
the difference in average halo mass between galaxies at the 84\%-ile and 16\%-ile of the colour distribution at a fixed redshift and surface brightness distribution is $0.00\pm0.11$~dex.

Using stellar mass instead of luminosity does not shift the inferred amplitude of correlations by more than the uncertainty.
We then conclude that our analysis did not find evidence for the existence of a correlation between halo mass and a secondary galaxy property at a fixed stellar mass.
Elliptical galaxies with a different size or colour for the same stellar mass have presumably had different assembly histories.
Our results indicate that, regardless of these differences, the average efficiency of massive elliptical galaxies in converting gas into stars, which is reflected by the stellar-to-halo mass ratio, is roughly constant as a function of size or colour.
The origin of the diversity in the size and colour distribution of these objects must be searched for by examining properties other than halo mass.

\begin{acknowledgements}

AS acknowledges funding from the European Union's Horizon 2020 research and innovation programme under grant agreement No 792916.
H. Hoekstra acknowledges support from Vici grant 639.043.512, financed by the Netherlands Organisation for Scientific Research (NWO).
MA acknowledges support from the European Research Council under grant number 647112.
MB is supported by the Polish National Science Center through grants no. 2020/38/E/ST9/00395, 2018/30/E/ST9/00698 and 2018/31/G/ST9/03388, and by the Polish Ministry of Science and Higher Education through grant DIR/WK/2018/12.
CH acknowledges support from the European Research Council under grant number 647112, and support from the Max Planck Society and the Alexander von Humboldt Foundation in the framework of the Max Planck-Humboldt Research Award endowed by the Federal Ministry of Education and Research.
H. Hildebrandt is supported by a Heisenberg grant of the Deutsche Forschungsgemeinschaft (Hi 1495/5-1) as well as an ERC Consolidator Grant (No. 770935).
KK acknowledges support from the Royal Society and Imperial College.
NRN acknowledges financial support from the “One hundred top talent program of Sun Yat-sen University” grant N. 71000-18841229.
AHW is supported by an European Research Council Consolidator Grant (No. 770935).

Based on data products from observations made with ESO Telescopes at the La Silla Paranal Observatory under program IDs 177.A-3016, 177.A-3017 and 177.A-3018, and on data products produced by Target/OmegaCEN, INAFOACN, INAF-OAPD and the KiDS production team, on behalf of the KiDS consortium. OmegaCEN and the KiDS production team acknowledge support by Deutsche Forschungsgemeinschaft, ERC, NOVA and NWO-M grants.

Funding for SDSS-III has been provided by the Alfred P. Sloan Foundation, the Participating Institutions, the National Science Foundation, and the U.S. Department of Energy Office of Science. The SDSS-III web site is http://www.sdss3.org/.

SDSS-III is managed by the Astrophysical Research Consortium for the Participating Institutions of the SDSS-III Collaboration including the University of Arizona, the Brazilian Participation Group, Brookhaven National Laboratory, Carnegie Mellon University, University of Florida, the French Participation Group, the German Participation Group, Harvard University, the Instituto de Astrofisica de Canarias, the Michigan State/Notre Dame/JINA Participation Group, Johns Hopkins University, Lawrence Berkeley National Laboratory, Max Planck Institute for Astrophysics, Max Planck Institute for Extraterrestrial Physics, New Mexico State University, New York University, Ohio State University, Pennsylvania State University, University of Portsmouth, Princeton University, the Spanish Participation Group, University of Tokyo, University of Utah, Vanderbilt University, University of Virginia, University of Washington, and Yale University.

{\em Author contribution:} All authors contributed to the development and writing of this paper. The authorship list is given in two groups: the lead authors (AS, CT, HHo), followed by an alphabetical group which includes those who have either made a significant contribution to the data products, or to the scientific analysis.

\end{acknowledgements}

\bibliographystyle{aa}
\bibliography{references}

\end{document}